\documentclass[10pt,final,journal,twocolumn]{IEEEtran}

\IEEEoverridecommandlockouts
\usepackage{color}

\usepackage{amsthm,amsmath,amssymb}
\usepackage{mathtools}

\usepackage{verbatim}

\usepackage{cite}
\usepackage{url}
\usepackage{cases}
\usepackage{stmaryrd}

\usepackage[inline]{enumitem}

\usepackage{graphicx}

\ifCLASSOPTIONcompsoc
  \usepackage[caption=false,font=normalsize,labelfont=sf,textfont=sf]{subfig}
\else
  \usepackage[caption=false,font=footnotesize]{subfig}
\fi

\usepackage[switch]{lineno}
\usepackage[named]{algo}
\usepackage[noend]{algpseudocode}
\usepackage{algorithm}

\usepackage[flushleft]{threeparttable} 
\usepackage{bm}
\newtheorem{theorem}{Theorem}

\newtheorem{lemma}[theorem]{Lemma}

\newtheorem{remark}{Remark}

\begin{document}
\title{Multi-IRS-Aided Doppler-Tolerant \\Wideband DFRC System}% Dual-Function Radar-Communications}

\author{Tong Wei,~\IEEEmembership{Student Member,~IEEE}, Linlong Wu,~\IEEEmembership{Member,~IEEE}, Kumar Vijay Mishra,~\IEEEmembership{Senior Member,~IEEE}, M. R. Bhavani Shankar,~\IEEEmembership{Senior Member,~IEEE}

%\IEEEmembership{Senior Member,~IEEE}
\thanks{This work was supported 
by Luxembourg National Research
Fund (FNR) through the SPRINGER Project under Grant C18/IS/12734677.}
\thanks{The authors are with the Interdisciplinary Centre for Security, Reliability and Trust (SnT), University of Luxembourg, Luxembourg City L-1855, Luxembourg. E-mail: \{tong.wei@, linlong.wu@, kumar.mishra@ext., bhavani.shankar@\}uni.lu.}
\thanks{The conference precursor of this paper appeared in the 2022 IEEE International Hybrid Symposium on Joint communications \& Sensing (JC\&S) \cite{tong2022multiple}.}
}

% make the title area
\maketitle

% As a general rule, do not put math, special symbols or citations
% in the abstract
\vspace{-1.2cm}
\begin{abstract}
Intelligent reflecting surface (IRS) is recognized as an enabler of future dual-function radar-communications (DFRC) by improving spectral efficiency, coverage, parameter estimation, and interference suppression. Prior studies on IRS-aided DFRC focus  either on narrowband processing, single-IRS deployment, static targets, non-clutter scenario, or on the under-utilized line-of-sight (LoS) and non-line-of-sight (NLoS) paths. In this paper, we address the aforementioned shortcomings by optimizing a wideband DFRC system comprising multiple IRSs and a dual-function base station that jointly processes the LoS and NLoS wideband multi-carrier signals to improve both the communications SINR and the radar SINR in the presence of a moving target and clutter. We formulate the transmit, {receive} and IRS beamformer design as the maximization of the worst-case radar signal-to-interference-plus-noise ratio (SINR) subject to transmit power and communications SINR. We tackle this nonconvex problem under the alternating optimization framework, where the subproblems are solved by a combination of Dinkelbach algorithm, consensus alternating direction method of multipliers, and Riemannian steepest decent. Our numerical experiments show that the proposed multi-IRS-aided wideband DFRC provides over $4$ dB radar SINR and $31.7$\% improvement in target detection over a single-IRS system.
\end{abstract}


\IEEEpeerreviewmaketitle

\section{Introduction}
Over the past few years,  intelligent reflecting surface (IRS) has emerged as a promising technology to achieve a smart wireless environment that allows enhanced coverage, security, and interference suppression \cite{renzo2020smart,huang2019reconfigurable}. An IRS comprises a large number of low-cost sub-wavelength passive meta-material elements, each of which is able to independently control the phase of the impinging signal and hence shape the radiation beampattern to alter the radio propagation environment \cite{renzo2020smart,wu2020towards,hodge2020intelligent}. The near passive behavior implies potential for large-scale IRS deployment without additional energy consumption when compared with the conventional relays \cite{garcia2020reconfigurable}. These characteristics of IRS have attracted considerable attention in both sensing \cite{aubry2021reconfigurable,zahra2021irs-aided} and communications communities \cite{lu2021intelligent,hodge2020intelligent}. % which have already studied the benefits of integrating IRS with several existing wireless applications, such as multiple-input-single-output (MIMO) radar, MIMO communications, and dual-function radar-communications (DFRC) \cite{irs_jrc_SecrecyRateOptMishra}.  to enhance the detection performance and the quality of service (QoS) with the stable data transmission \cite{buzzi2021radar,wu2020towards,liu2022joint}.

Initial investigations of IRS were limited to wireless communications to enhance, for instance, the coverage, spectral efficiency, energy saving, secrecy rate, and interference suppression; see e.g. \cite{wu2020towards} and references therein. %\cite{zhou2021joint,abeywickrama2020intelligent,najafi2021physics,elzanaty2021reconfigurable,xu2020resource,yu2020robust,cao2021outage,mei2020joint,saeidi2021weighted,zheng2021double,li2020weighted,mei2021cooperative,wu2019intelligent,Ur2021joint,wu2020joint,tang2021wireless}. 
These studies utilized IRS to compensate for the end-to-end (transmitter-IRS-receiver) path loss \cite{xu2020resource,tang2021wireless}. Further, the base station (BS) and reflecting surfaces employed, respectively, active and largely passive beamformers \cite{Ur2021joint,wu2020joint}, which could be designed jointly. For example, the IRS-aided channel in \cite{zhou2021joint} is estimated for each reflecting element using pilot symbols followed by a transmit precoder design. In \cite{abeywickrama2020intelligent}, IRS was employed to minimize the total transmit power while guaranteeing the signal-to-interference-plus-noise ratio (SINR) among all users. %that reveals the relation between the reflection amplitude and phase-shift, is proposed to facilitate the implementation. 
Some recent studies \cite{elzanaty2021reconfigurable} employ IRS to correctly estimate the position and orientation of the mobile user to improve the quality-of-service (QoS). The system model in these investigations employed a single IRS thereby limiting the flexibility of deployment and the degrees-of-freedom for resolving the large-scale channel fading. 

Multiple IRSs, if optimally deployed, have the potential to exploit the multiplicative beamforming gain to further enhance the QoS \cite{yu2020robust,zheng2021double,mei2021cooperative}. For example, in secure communications \cite{yu2020robust}, IRS is useful in establishing a favorable propagation environment, wherein the sum rate of legitimate users is maximized under the constraint on leakage of information to the potential eavesdroppers. Note that \cite{yu2020robust} ignores the interaction between the IRSs, thus simplifying the design and analysis. When the line-of-sight (LoS) path between the BS and users is blocked, the double IRS paradigm has been shown to yield the virtual LoS in multi-user multiple-input multiple-output (MIMO) communications \cite{zheng2021double}. A multi-IRS-aided wireless communications system developed in \cite{mei2021cooperative} leverages short-distance LoS channel between two adjacent IRSs to receive the transmit message via multiple reflections. These multi-IRS studies  limit their analyses to only the non-line-of-sight (NLoS) path between the transmitter and receiver via IRS.

%-----------------------------------------------------------------------------------
\begin{table*}
\centering
\caption{Comparison with the state-of-the-art}
\vspace{-0.3cm}
\label{tbl:priorcomp}
\scriptsize
\begin{threeparttable}
\begin{tabular}{l|c|c|c|c|c|c|l}
\hline cf. & Radar & Communications & IRS; Phases & Targets; Clutter & Radar paths\tnote{a}    & Users & Design metric  \\
\hline 
\hline 
\cite{he2022ris}            & PA\tnote{b} & SC-SISO\tnote{c} & Double; Continuous & Static, multiple; Yes & LoS, NLoS & Single  & Communications SINR  \\
\hline 
\cite{jiang2022intelligent}  & MIMO & SC-MIMO\tnote{d} & Single; Continuous & Static, single; No & LoS, NLoS  & Single & Radar SNR \\
\hline 
\cite{wang2021ris}           & MIMO & SC-MIMO & Single; Continuous & Static, single; No & LoS, NLoS & Multiple  & Detection probability \\
\hline 
\cite{liu2022joint}          & MIMO & SC-MIMO  & Single; Continuous & Static, single; Yes & LoS, NLoS & Multiple  & Radar SINR \\
\hline 
\cite{song2022joint}  & {MIMO} & {SC-MISO}  & {Single; Continuous} & {Static, multiple; No} & {NLoS} & {Single}  & {Minimum radar gain} \\
\hline 
\cite{zhu2022intelligent}    & MIMO & SC-MIMO  & Single; Continuous & Static, single; Yes & LoS & Multiple  & Sum-rate\\
\hline 
\cite{hua2022joint}          & MIMO & SC-MIMO & Single; Continuous & Static, multiple; No & LoS & Multiple  & Transmit power \\
\hline 
\cite{wang2021joint}        & MIMO & SC-MIMO & Single; Continuous & Static, multiple; No & LoS & Multiple  & Beampattern, MUI  \\
\hline 
\cite{wang2022jointwaveform} & MIMO & SC-MIMO  & Single; Discrete & Static, multiple; No & LoS & Multiple  & Beampattern, MUI  \\
\hline
\cite{esmaeilbeig2023quantized} & {MIMO} & {SC-MIMO}  & {Single; Discrete} & {Static, single; No} & {NLoS} & {Multiple}  & {Tradeoff design}   \\
\hline
This paper & W-MIMO\tnote{e} & MIMO-OFDM & Multiple; Continuous & Moving, single; Yes & LoS, NLoS & Multiple & Minimum radar SINR \\
\hline
\end{tabular}
    \begin{tablenotes}[para]
	\item[a]The communications system utilizes the LoS and NLoS paths in all these works.
	\item[b] PA: phased-array.
	\item[c] SC-SISO: single-carrier single-input-single-output.
	\item[d] SC-MIMO: single-carrier MIMO.
	\item[e] W-MIMO: wideband MIMO.
	\end{tablenotes}
    \end{threeparttable}
    \normalsize
    \vspace{-1em}
\end{table*}

%--------------------------------------------------------------------

%\vspace{-0.4cm}
The investigations into the potential of IRS towards enhancing the performance of, primarily, MIMO radar followed the spurt of IRS research in communications \cite{lu2021intelligent,aubry2021reconfigurable,buzzi2022foundations,wang2022joint,zahra2021irs-aided,aubry2021ris}.
The focus of these works has been similar to IRS-aided communications, i.e. use IRS to aid the radar in detecting NLoS targets. There is a rich heritage of research on non-IRS-based NLoS radar; see e.g. \cite{watson2019non,wei2022nonline} and references therein. However, these systems require accurate knowledge of the environment and geometry (such as walls and buildings) {\em apriori}. Further, unlike an IRS-assisted system, they are unable to alter and control the wireless media. This makes IRS very attractive for remote sensing of hidden or blocked objects. In \cite{aubry2021reconfigurable}, IRS facilitated coverage extension of an NLoS radar. %Meanwhile, the achievable signal-to-noise ratio (SNR) in a different area is derived via the radar equation. 
This was extended to monostatic/ bistatic and LoS/ NLoS radar systems in \cite{buzzi2022foundations} and the receive signal-to-noise ratio (SNR) was maximized by optimizing the phase-shifts. This study revealed that, when the IRS is far away from the radar transmitter or receiver, the detection performance has only marginal gain. A very recent study in \cite{wang2022joint} considered multiple IRSs to enhance the transmit power toward target-of-interest with constraints on the clutter backscatter. % minimum illumination power at the target directions while keeping the clutter scatterers radiation and total transmit power less than the threshold. %The resulting maximin problem is solved by alternating maximization (AM) framework, in which the semidefinite relaxation (SDR) method is utilized to tackle the subproblems.   

Lately, there has been significant interest in characterizing IRS performance for integrated sensing and communications (ISAC) systems \cite{wang2021joint,sankar2021joint,wang2022jointwaveform,hua2022joint,zhu2022intelligent,sankar2022beamforming,liu2022joint2}. The motivation for developing ISAC systems lies in addressing the increasing spectrum congestion by designing common hardware and waveforms for both radar and communications \cite{mishra2019toward,wu2022resource,liu2020joint,luo2022joint}. % \cite{mishra2019toward,dokhanchi2019a,wu2022resource,liu2020joint,alaee2020information,qian2018joint,he2020joint,wei2022joint,wang2021ris,hassanien2016dual,wang2019dual,yang2020dual,cheng2021transmit,cheng2021hybrid,xu2021a,johnston2022mimo}.
%Based on whether sharing the transmit platform, most of the current ISAC systems can be divided into two categories, i.e., the coexistence of radar-communications (CRC) and DFRC. For the coexistence scheme, the radar and communications operate independently in the same frequency with their own set of separate waveforms, which inevitably leads to  mutual interference (MI) \cite{qian2018joint,he2020joint,wei2022joint,wang2021ris}.
These dual-function radar-communications (DFRC) units \cite{hassanien2016dual} %is developed to share the transmitter for both sensing and communications which does not exist the MI and shows the competitive 
have the advantages of resource sharing, hardware cost and energy efficiency. %For example, in \cite{hassanien2016dual}, the data symbol is embedded in the radar waveform and the communications function can be achieved by adjusting the sidelobe level. To reduce the energy consumption, in \cite{wang2019dual}, the sparse array configuration, which requires a few radio frequency chains, are developed for the DFRC system. Moreover, in \cite{cheng2021transmit}, the low-resolution DACs are utilized to reduce the hardware complexity further. In \cite{cheng2021hybrid}, the hybrid beamforming technique is extended to wideband DFRC, in which both the transmit beampattern of radar and the spectral efficiency of the user are optimized simultaneously.
%Based on the previous discussion, IRS is able to achieve smart wireless propagation and hence is appropriate to integrate into the DFRC system  With regard to this scenario, in \cite{jiang2022intelligent}, 
A single-IRS-aided DFRC proposed in \cite{jiang2022intelligent} was radar-centric in that it maximized the radar SNR while utilizing the reflecting surface to simultaneously facilitate the target detection and single-user communications. %Meanwhile,  the radar SNR is maximized while ensuring the communications SNR which is known as the radar-centric design. 
This was extended to multiple users in \cite{wang2021joint}, wherein the radar transmit beampattern was synthesized and was followed by a minimization of multi-user interference (MUI) to guarantee the communications QoS. {On the other hand, \cite{wang2021ris} used IRS to enhance the radar detection probability and ensure certain communications SINR over all users. To improve the robustness of IRS-aided ISAC, \cite{song2022joint} maximized the minimum gain towards all the targets while guaranteeing the communications SNR.}
While several theoretical works assume the IRS phase-shifts to be continuous-valued, in practice, the shifts only admit discrete/ quantized values. This aspect was analyzed in the IRS-aided DFRC system suggested in \cite{wang2022jointwaveform}. {The semi-active IRS in \cite{esmaeilbeig2023quantized} employed discrete phase-shifts and simultaneously enhanced the radar SNR and communications SNR.}  
A few other recent studies on communications-centric DFRC design, where the IRS facilitates in maximizing secrecy rates \cite{irs_jrc_SecrecyRateOptMishra}. {Recently, in \cite{luo2022joint}, IRS is utilized to enhance the communications sum-rate while guarantee the radar detection performance via shaping the transmit beampattern of BS. Compared with \cite{luo2022joint} which deploys IRS only for communications user, the proposed method utilize IRS to boost the overall performance of DFRC system.}
% the discrete IRS-aided DFRC is considered to minimize the MUI and Cramer-Rao bound for direction-of-arrivals (DoAs) estimation, simultaneously. It is seen that IRS can also be utilized in the CRC to improve the communications SINR and guarantee the radar SINR \cite{he2022ris}. 

{Nearly all of the aforementioned single-IRS-assisted DFRC approaches focus on a stationary target and narrowband transmission. While the stationary target is not a very practical assumption from the radar perspective, the narrowband setting limits the application to lower frequency bands which is incompatible with the existing push towards mmWave or THz system. Further, the single IRS formulations also limits the field of view and multiple IRS settings remove any such restriction offering a full view of the scene.}
%Further, single IRS formulations limit the coverage region and ignore the spatial diversity of target radar cross-section (RCS). %To address these shortcomings of prior research, 
Towards this end, in this paper, we propose the utilization of multiple IRSs to assist a wideband DRFC system. In particular, we employ orthogonal frequency-division multiplexing (OFDM) waveform to detect a moving target and communicate with multiple users simultaneously. We devise a Doppler filter bank against an unknown Doppler shift at the radar receiver. We show that by the proper design of the transmit beamforming, phase-shift matrix, and Doppler filter-banks, we maximize the average radar SINR over all subcarriers while ensuring that the average SINR among all users is greater than a predetermined threshold, thus guaranteeing the communications QoS. %To the best of our knowledge, the multi-IRS-aided wideband DFRC has not been well studied in the existing literature yet. 
Table~\ref{tbl:priorcomp} summarizes the key differences between our work and some major closely-related prior studies.

Preliminary results of this work appeared in our conference publication \cite{tong2022multiple}, where we introduced wideband IRS-assisted DFRC but ignored the LoS paths for both sensing and communications, did not consider moving target, and omitted detailed performance evaluations. In this work, we include these critical assumptions and our main contributions are:\\
\textbf{1) Wideband IRS-aided DFRC:} We propose a  comprehensive multi-IRS-aided wideband OFDM-DFRC model, which includes moving target in the radar scene and multiple single-antenna users for communications. We also consider both LoS and NLoS paths for radar and communications. The wideband design allows for varying beamformer weights with respect to subcarrier frequencies and thereby offset the beam squint effect  \cite{liu2010wideband,ma2021wideband}. Our proposed system, therefore, subsumes current stationary target, narrowband, non-IRS, or single-IRS DFRC studies \cite{liu2022joint,jiang2022intelligent,cheng2021transmit}; also, see Table~\ref{tbl:priorcomp}. \\  
\textbf{2) Doppler-tolerant IRS-DFRC {model}:} %Due to the fast-moving target, the Doppler shift, which is unknown in advance, can inevitably lead the performance loss. Toward this end, 
Analogous to a Doppler filter bank in a conventional radar receiver, we design multiple receive filters to maximize the worst-case radar SINR accounting for all possible Doppler slices. This robust design is subjected to the constarints of subcarrier transmit power and the minimum communications SINR among all users. We then solve the resulting nonconvex maxmin problem, which involves fractional quartic objective function accompanied by the difference of convex (DC) and unimodular constraints. \\
\textbf{3) Alternating optimization (AO) for joint design:} We develop an alternating maximization (AM) framework to tackle the above nonconvex problem. % with the suboptimal solution. 
We first utilize the semidefinite relaxation (SDR) to obtain the closed-form solution for the Doppler filter bank design. Then, we combine the Dinkelbach and majorization method to tackle the transmit beamformer design subproblem. Finally, the consensus alternating direction of multipliers (C-ADMM) \cite{cheng2021hybrid} and Riemannian steepest decent (RSD) \cite{wang2021joint} approaches are jointly used to approximately solve the subproblem of phase-shift design. \\
\textbf{4) Extensive performance evaluation:} We validate our model and methods through comprehensive numerical experiments. %are provided to illustrate the superior performance of the proposed algorithm under different simulation setups. As expected, 
Our proposed method achieves an enhanced radar SINR over its non-IRS, single-IRS, and narrowband counterparts. Our theoretical analyses and experimental investigation of the proposed IRS-aided DFRC reveal a trade-off between the communications and radar performance. 

The remainder of this paper is organized as follows. In the next section, we introduce the signal model and problem formulation for multi-IRS-aided wideband DFRC system. % are first introduced in Section \uppercase\expandafter{\romannumeral2}. Next, 
In Section \uppercase\expandafter{\romannumeral3}, we develop our AM-based algorithm to tackle the formulated optimization problem, in which the corresponding subproblems are solved iteratively. We evaluate our methods in Section \uppercase\expandafter{\romannumeral4} through extensive numerical examples. We conclude in %are utilized to demonstrate the effectiveness and superiority of the proposed method. Finally, 
Section \uppercase\expandafter{\romannumeral5}. % summarizes this work.

\emph{Notations:} 
Throughout this paper, vectors and matrices are denoted by lower case boldface letter and upper case boldface letter, respectively. The notations $(\cdot)^T$, $(\cdot)^{\ast}$ and $(\cdot)^H$ denote the operations of transpose, conjugate, and Hermitian transpose, respectively; ${\bf I}_L$ and ${\bf 1}_L$ denote the $L\times L$ identity matrix and all-ones vector of length $L$, respectively; {$\odot$ and} $\otimes$ are the {Hadamard and} Kronecker product, {respectively}; $\Re(\cdot)$ and $\Im(\cdot)$ represent the real and imaginary parts of a complex number, respectively; $\mathrm{vec}(\cdot)$ is the vectorization of its matrix argument; $\mathrm{diag}(\cdot)$ and $\mathrm{blkdiag}(\cdot)$ denote the diagonal and block diagonal matrix, respectively; $|\cdot|$, $\|\cdot\|_2$ and $\|\cdot\|_F$ represent the magnitude, $\ell_2$-norm, and Frobenius-norm, respectively; $(\cdot)^{(n)}$ denotes the value of the variable at the $n$-th outer iteration; and {$\nabla_E(\cdot)$ and $\nabla_R(\cdot)$ are the Euclidean and Riemannian gradient operator, respectively.} 

\section{System Model and Problem Formulation}

\begin{figure}[!t]
\centering
\includegraphics[width=0.35\textwidth]{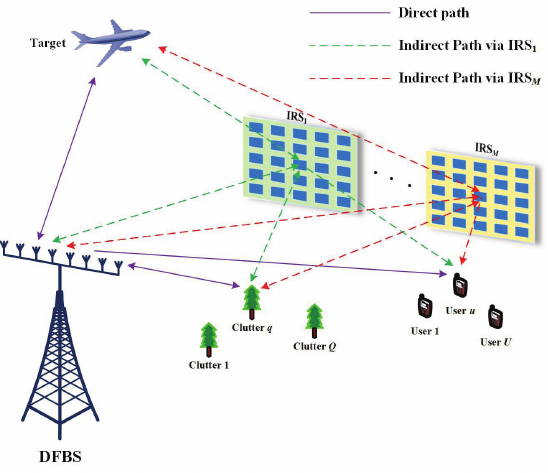}
\vspace{-0.1cm}\caption{Simplified illustration of a multi-IRS-aided wideband DFRC system.}
\label{fig:system}
\vspace{-0.5cm}
\end{figure}

Consider a multi-IRS-aided wideband OFDM-DFRC system consisting of a dual-function transmitter, a colocated radar receiver, and $M$ IRSs (Fig.~\ref{fig:system}). The dual-function transmitter and the radar receiver are closely deployed with the dual-function base station (DFBS). We denote the reflecting elements of $m$-th IRS and the number of antennas at the DFBS transmitter and the receiver by {$N_{I_m}$, $N_{B_t}$ and $N_{B_r}$}, respectively.  The uniform inter-element spacing for the {$m$-th IRS}, DFBS transmit and receive arrays are {$d_{I_m}$, $d_{B_t}$ and $d_{B_r}$}, respectively. The IRS-aided DFRC system aims to detect a moving target in the presence of $Q$ clutter {scatterers} while simultaneously serving $U$ downlink (DL) single-antenna users in a two-dimensional (2-D) Cartesian plane. Assume that the BS, radar target, $q$-th clutter patch, $m$-th IRS and are located respectively at the coordinates, {${\bf p}_B=[x_B,y_B]$, ${\bf p}_T=[x_T,y_T]$, ${\bf p}_{C_q}=[x_C(q),y_C(q)]$ and ${\bf p}_{I_m}=[x_I(m),y_I(m)]$. To simplify the notation, hereafter, we use the subscript $B_t$, $B_r$, $C_q$ and $I_m$ denote the DFBS transmitter, radar receiver, $q$-th clutter and $m$-th IRS.  Further, the subscript $B$, and $T$, denote the DFBS and target respectively}.
%
%The OFDM transmit signal has $K$ subcarriers. In this paper, we consider both the direct and indirect (i.e., LoS and NLoS) links for radar and communications. T
%Then, we denote the various LoS/NLoS paths for radar and communications as: \\
%\emph{Radar}: DFBS-target-DFBS (or direct path), DFBS-IRS-target-DFBS,  DFBS-target-IRS-DFBS, DFBS-IRS-target-IRS-DFBS;\\
%\emph{Communications}: DFBS-users (or direct path) and DFBS-IRS-uesrs.
%
\vspace*{-0.1in}
\subsection{Transmit signal}
The normalized transmit data symbol at the $k$-th subcarrier is ${\bf s}_k=[s_{k,1},\cdots,s_{k,U}]^T\in\mathbb{C}^{U\times 1}$, where $k=1,\cdots,K$ and $\mathbb{E}\{{\bf s}_k{\bf s}_k^H\}={\bf I}_U$. Meanwhile, the data symbol is modulated and spread over $K$ OFDM subcarriers. In the wideband OFDM system, the transmit steering vectors are frequency dependent and change with the subcarrier frequencies; this leads to the beam-squint effect \cite{ma2021wideband}. To mitigate this, we utilize a frequency-dependent beamforming technique to preprocess the transmit data symbol ${\bf s}_k$ in the frequency-domain. For the $k$-th subcarrier, denote the digital beamforming matrix by ${\bf F}_k=[{\bf f}_{k,1},\cdots,{\bf f}_{k,U}]\in\mathbb{C}^{N_{B_t}\times U}$, then the transmitted signal becomes ${\bf F}_k{\bf s}_k\in\mathbb{C}^{N_{B_t}\times1}$. Applying the $K$-point inverse fast Fourier transform (IFFT) to each element of the frequency-domain signal ${\bf F}_k{\bf s}_k$, the transmit baseband signal at time instant $t$ is
\begin{equation}\label{1}
  {\bf x}_{B_t}(t)=[x_1(t),\cdots,x_{N_{B_t}}(t)]^T=\sum_{k=1}^{K}{\bf F}_k{\bf s}_k e^{\mathrm{j}2\pi{f_k}t},
\end{equation}\normalsize
where $t\in(0,\triangle t]$ and $\triangle t$ denotes the OFDM duration excluding the cyclic prefix (CP) with length $D$, $f_k=(k-1)\triangle f$ denotes the baseband frequency at the $k$-th subcarrier and $\triangle f$ is the subcarrier spacing. To guarantee the orthogonality of the subcarriers, the frequency spacing is set to ${\triangle f}=1/{\triangle t}$. Meanwhile, for wideband DFRC, the total transmit power should meet the system requirement. In order to fully utilize the bandwidth, herein, we assume the transmit power satisfies 
$\|{\bf F}_k\|_F^2 \leq \mathcal{P}_k, k=1,\cdots,K$, 
where $\mathcal{P}_k$ denotes the maximum transmit power assigned to the $k$-th subcarrier. The baseband signal is then upconverted resulting in the transmission of $\mathbf{x}(t) = \mathbf{x}_{B_t}(t)e^{\mathrm{j}2\pi f_ct}$, where $f_c$ denotes the carrier frequency.
\subsection{Channel and operating conditions}
For radar system, we assume the transmit signal impinges on the target located at ${\bf p}_T$ moving with a  velocity ${\bf v}=[v_x,v_y]$, where $v_x$ and $v_y$ are the velocity components along the $x$ and $y$-axes, respectively. The transmit signal is then reflected to the radar receiver from both the direct and indirect paths. Here, we consider the echoes arising via the following paths: Tx-target-Rx (\textbf{path $1$} or the \textbf{direct path}), Tx-IRS-target-Rx (\textbf{path $2$}), Tx-target-IRS-Rx (\textbf{path $3$}), and Tx-IRS-target-IRS-Rx (\textbf{path $4$})
\footnote{
{{
First, some blockages (e.g., buildings) in urban areas result in a weaker direct path between DFBS and target/user, which indicates the prominent role of the IRS-involved indirect paths. Therefore, we consider both direct and indirect paths and unify them in a single system model. Further, for \textbf{path} 4, we omit the case with the different forward and backward IRSs because of lower received power.
}
}}.  
{To detail the channel model, we first define the frequency-dependent steering vector of the transmitter, receiver and IRS as 
\begin{subequations} \label{ste_vec}
 \begin{align} 
 &{\bf a}_{B_t}(\theta,f_k)\!=\! [ 1,\!e^{-\mathrm{j}v_t(\theta,f_k)}, \!\cdots\!,\!e^{-\mathrm{j}(N_{B_t}-1)v_{B_t}(\theta,f_k)}]^T, \\
 &{\bf a}_{B_r}(\theta,f_k)\!=\! [ 1,\!e^{-\mathrm{j}v_r(\theta,f_k)}, \!\cdots\!,\!e^{-\mathrm{j}(N_{B_r}-1)v_{B_r}(\theta,f_k)}]^T, \\
 &{\bf a}_{I_m}(\theta,f_k)\!=\! [ 1,\!e^{-\mathrm{j}v_r(\theta,f_k)}, \!\cdots\!,\!e^{-\mathrm{j}(N_{I_m}-1)v_{I_m}(\theta,f_k)}]^T
 \end{align}
\end{subequations}\normalsize
where $v_{\nu}(\theta,f_k)=2{\pi}(f_k+f_c)(\frac{d_{\nu}\sin(\theta)}{c}), \nu\in \{B_t,B_r,I_m\}$ denotes the  spatial-frequency shift.}

Denote the  channel coefficient of radar direct path that includes target reflectivity and distance-dependent path loss by {$\alpha_{1}=\sqrt{{\alpha_T}K_0(\frac{r_0}{2r_{BT}})^{\epsilon_{1}}}$, where ${\alpha_T}$ is the target RCS, $K_0$ is the signal attenuation at the reference distance $r_0$, $r_{BT}=\|{\bf p}_T-{\bf p}_B\|_2$ is the distance between DFBS and target, $\epsilon_{1}$ denotes the direct path loss exponent}. The Doppler frequency of the target with respect to the direct path is $f_{D,1}$. The target angle of arrival/ departure (AoA/AoD) with respect to DFBS is {$\theta_{BT}$}\footnote{Hereafter, all the AoAs/ AoDs are measured with respect to the array broadside direction and  positive when moving clockwise.}. The propagation delay for the path between the DFBS and the target, the differential propagation delay for the $n_{B_t}$-th transmitter with respect to the reference (first) transmitter, and the differential propagation delay for the $n_{B_r}$-th receiver with respect to the reference (first) receiver are, respectively,
\begin{align}\label{eq:delay}
  &{\tau_{BT}  =\frac{\|{\bf p}_T-{\bf p}_B\|_2}{c}}, ~
  \tau_{n_{B_t}}  =\frac{(n_{B_t}-1)d_{B_t}\sin\theta_{BT}}{c}, \nonumber\\ &\tau_{n_{B_r}} =\frac{(n_{B_r}-1)d_{B_r}\sin\theta_{BT}}{c},
\end{align} \normalsize
where $c=3\times10^8$ m/s is the speed of light. Hence, the total time delay from the $n_{B_t}$-th transmitter and $n_{B_r}$-th receiver via \textit{direct path} is
\begin{align}
\tau_{\textrm{dir}, n_{B_r}, n_{B_t}}=2\tau_{BT}+\tau_{n_{B_t}}+\tau_{n_{B_r}}.   
\end{align}\normalsize
Meanwhile, the propagation delays from the DFBS to $m$-th IRS and target to $m$-th IRS are
 $\tau_{BI_m}  =\frac{\|{\bf p}_B-{\bf p}_{I_m}\|_2}{c}$ and 
 $\tau_{TI_m}  =\frac{\|{\bf p}_T-{\bf p}_{I_m}\|_2}{c}$, respectively.
{Thus, the total delay of indirect  path $2$, $3$ and $4$ from the $n_{B_t}$-th DFBS transmit antenna to the $n_{B_r}$-th DFBS receive antenna through the $n_{I_m}$-th element of the $m$-th IRS  
are  $\tau_{\textrm{ind}_2, n_{I_m}, n_{B_t}, n_{B_r}}=\tau_{BT}+\tau_{BI_m}+\tau_{TI_m}$, $\tau_{\textrm{ind}_3, n_{I_m}, n_{B_t}, n_{B_r}}=\tau_{BT}+\tau_{BI_m}+\tau_{TI_m}$ and $\tau_{\textrm{ind}_4, n_{I_m}, n_{B_t}, n_{B_r}}=2\tau_{BI_m}+2\tau_{TI_m}$.} 
This includes the differential path delays induced by the uniform linear array configuration of the DFBS and IRS. 

Then, the Doppler frequencies with respect to the paths $1$, $2$, $3$, and $4$ %Tx-target-Rx (\textbf{path $1$} or the \textbf{direct path}), Tx-IRS-target-Rx (\textbf{path $2$}), Tx-target-IRS-Rx (\textbf{path $3$}), and Tx-IRS-target-IRS-Rx (\textbf{path $4$}) 
are, respectively, \cite{sun2019target} 
\begin{subequations} \label{dop_shift}
\begin{align}
   & f_{D_1} =\frac{f_c}{c}\left(\frac{\langle{\bf v},{\bf p}_T-{\bf p}_B\rangle}{\|{\bf p}_T-{\bf p}_B\|_2}+\frac{\langle{\bf v},{\bf p}_T-{\bf p}_B\rangle}{\|{\bf p}_T-{\bf p}_B\|_2}\right), \label{dop1}\\
  & f_{D_2,I_m}\!=\!\frac{f_c}{c}\left(\frac{\langle{\bf v},{\bf p}_T-{\bf p}_B\rangle}{\|{\bf p}_T-{\bf p}_B\|_2}\!+\!\frac{\langle{\bf v},{\bf p}_T-{\bf p}_{I_m}\rangle}{\|{\bf p}_T-{\bf p}_{I_m}\|_2}\right), \forall m,\label{dop2}\\
  & f_{D_3,I_m}  \!=\!\frac{f_c}{c}\left(\frac{\langle{\bf v},{\bf p}_T-{\bf p}_B\rangle}{\|{\bf p}_T-{\bf p}_B\|_2}\!+\!\frac{\langle{\bf v},{\bf p}_T-{\bf p}_{I_m}\rangle}{\|{\bf p}_T-{\bf p}_{I_m}\|_2}\right), \forall m, \label{dop3} \\
  & f_{D_4,I_m}  \!=\!\frac{f_c}{c}\left(\frac{\langle{\bf v},{\bf p}_T-{\bf p}_{I_m}\rangle}{\|{\bf p}_T-{\bf p}_{I_m}\|_2}\!+\!\frac{\langle{\bf v},{\bf p}_T-{\bf p}_{I_m}\rangle}{\|{\bf p}_T-{\bf p}_{I_m}\|_2}\right), \forall m.  \label{dop4} 
\end{align}
\end{subequations}\normalsize 

%and 
%\begin{align}\label{dop4}
%  & f_{D_4,I_m}  =\frac{f_c}{c}\left(\frac{\langle{\bf v},{\bf p}_B-{\bf p}_{I_m}\rangle}{\|{\bf p}_B-{\bf p}_{I_m}\|_2}+\frac{\langle{\bf v},{\bf p}_B-{\bf p}_{I_m}\rangle}{\|{\bf p}_B-{\bf p}_{I_m}\|_2}\right), \; \forall m.
%\end{align}\normalsize

We make the following assumptions about the IRS-aided DFRC and channel parameters:
\begin{description}
\item[A1] ``Known channel state information (CSI)'': The CSI matrix from the transmitter (or IRSs) to users are estimated in advance \cite{he2021channel}. 
\item[A2] ``Bandwidth-invariant Doppler'':  The bandwidth of OFDM signal is much smaller than the carrier frequency, i.e., $K \triangle f\ll f_c$. Hence, the phase-shifts arising from the Doppler effect are identical over all subcarriers. 
\item[A3] ``Constant Doppler shifts'': The Doppler frequency of target does not change during each OFDM symbol duration $K \triangle t$, i.e., ${dv_t}/{dt}\ll {c}/{(2f_c (  \triangle t)^2)}$.

\item[A4] ``Constant IRS phase-shifts'': {Because IRS does not contain the baseband signal processing unit and hence is a kind of narrowband device,} the phase-shift of IRS is constant at different subcarriers: ${\bf\Phi}_m(f_k) = {\bf\Phi}_m, m=1,\cdots,M$ over all subcarrier frequencies $f_k$. 
\end{description}
\subsection{Radar receiver}
The received signal from the target at the $n_{B_r}$-th radar antenna via \textbf{direct path} is a delayed, modulated, and scaled version of the transmit signal \eqref{1}, that is,
\begin{align}\label{3}
 s_{\textrm{dir},{n_{B_r}}}(t) = &\sum_{n_t=1}^{N_t}\alpha_{1}x_{n_t}(t-\tau_{\textrm{dir},n_{B_r},n_{B_t}}) \nonumber\\
 &e^{\mathrm{j}2\pi{f_c}(t-\tau_{\textrm{dir},n_{B_r},n_{B_t}})}  
 e^{\mathrm{j}2\pi{f_{D,1}}(t-\tau_{\textrm{dir},n_{B_r},n_{B_t}})}.
\end{align}\normalsize
Stacking the echoes for all receive antennas and removing carrier frequency $e^{\mathrm{j}2\pi{f_c}t}$, the $N_{B_r} \times 1$ baseband signal vector is 
\begin{align}\label{5}
  {\bf s}_{\textrm{dir}}(t) &= [s_{\textrm{dir},1}(t),\cdots,s_{\textrm{dir},{N_{B_r}}}(t)]^T \nonumber\\
  =&\sum_{k=1}^{K} \widetilde{\alpha}_{1,k}{\bf a}_{B_r}(\theta_{BT},f_k){\bf a}_{B_t}^T(\theta_{BT},f_k){\bf F}_k{\bf s}_ke^{\mathrm{j}2\pi{f_k}t},
\end{align}\normalsize
where {$\widetilde{\alpha}_{1,k}\!=\!{\alpha}_{1}e^{\mathrm{j}2\pi{f_{D,1}}\triangle t}
e^{-\mathrm{j}2\pi({f_c}+{f_k})2\tau_{BT}}$} denotes the complex-value channel gain in terms of RCS, path loss, Doppler shift and propagation delay\footnote{Herein, the complex-valued term $e^{-\mathrm{j}2\pi{f_{D,1}}\tau_{d}}$ is omitted because $f_{D,1}\tau_{d}=const./c^2\approx0$.}, {${\bf a}_{B_t}(\theta,f_k)\in\mathbb{C}^{N_{B_t}\times 1}$ and ${\bf a}_{B_r}(\theta,f_k)\in\mathbb{C}^{N_{B_r}\times 1}$ are the space-frequency steering vector of the dual-function transmitter and radar receiver expressed as \eqref{ste_vec}.}
%, respectively, as 
%\begin{align}
% &{\bf a}_t(\theta,f_k)\!=\! [ 1,\!e^{-\mathrm{j}v_t(\theta,f_k)}, \!\cdots\!,\!e^{-\mathrm{j}(N_t-1)v_t(\theta,f_k)}]^T, 
% {\bf a}_r(\theta,f_k)\!=\! [ 1,\!e^{-\mathrm{j}v_r(\theta,f_k)}, \!\cdots\!,\!e^{-\mathrm{j}(N_r-1)v_r(\theta,f_k)}]^T, 
%\end{align}\normalsize
%where $v_{t,r}(\theta,f_k)=2{\pi}(f_k+f_c)(\frac{d_{t,r}\sin(\theta)}{c})$ denote the spatial-frequency shift.
Sampling \eqref{5} at the rate $K/\triangle t$ within the symbol duration yields the $N_{B_r} \times 1$ vector 
\begin{align}\label{dis}
  {\mathbf s}_{\textrm{dir}}[\widetilde{n}]&= [s_{\textrm{dir},1}[\widetilde{n}],\cdots,s_{\textrm{dir},{N_{B_r}}}[\widetilde{n}]]^T \nonumber \\
  = \sum_{k=1}^{K}&\widetilde{\alpha}_{1,k}{\bf a}_{B_r}(\theta_{BT},f_k){\bf a}_{B_t}^T(\theta_{BT},f_k){\bf F}_k{\bf s}_ke^{\mathrm{j}2\pi{f_k}\frac{\widetilde{n}\triangle t}{K}}, 
\end{align}\normalsize
where $\widetilde{n}=1,\cdots,K$ denotes the discrete-time sample index. Combining all $K$ samples, {we express \eqref{dis} in the matrix form as} 
\begin{align}\label{dis_m}
  {\bf S}_{\textrm{dir}}= [{\mathbf s}_{\textrm{dir}}[1],\cdots,{\mathbf s}_{\textrm{dir}}[K]]\in\mathbb{C}^{N_r\times K}. 
\end{align}\normalsize
Applying $K$-point FFT to each row of \eqref{dis_m} yields the $N_{B_r}\times 1$ frequency-domain vector for the $k$-th subcarrier as 
\begin{equation}\label{6}
     { \widetilde{\bf s}}_{\textrm{dir}}[f_k]={\bf A}_{\textrm{dir},k}{\bf F}_k{\bf s}_k,  
\end{equation}\normalsize
where {${\bf A}_{\textrm{dir},k}\!=\!\widetilde{\alpha}_{1,k}{\bf a}_{B_r}(\theta_{BT},f_k){\bf a}_{B_t}^T(\theta_{BT},f_k)$} is the direct channel response on the $k$-th subcarrier. 

For the indirect path, note that IRS is modeled as a linear array in a 2-D plane. This can be easily extended to 3-D geometry by modifying the corresponding steering vector \cite{sankar2021joint}. For the indirect path $2$, i.e., the path traversing DFBS-IRS-Target-DFBS, we use the distance-dependent path loss model again as {$\alpha_{I_m, 2}=\sqrt{{\alpha_T}K_0(\frac{r_0}{r_{BT}+r_{TI_m}})^{\epsilon_{2}}}$, where $\epsilon_{2}$ denotes the corresponding path loss exponent.}  The received signal of $n_{B_r}$-th element for path $2$ {via $m$-th IRS} is
\begin{align}
 s_{{\textrm{ind}_2,I_m,n_{B_r}}(t)} = & \sum_{n_{B_t}=1}^{N_{B_t}}\alpha_{I_m, 2} \sum_{n_{I_m}=1}^{N_{I_m}}x_{n_{B_t}}\left(t-\hat\tau_{\textrm{ind}_2}\right) \nonumber\\
 & e^{\mathrm{j}2\pi{f_c}\left(t-\hat\tau_{\textrm{ind}_2}\right)}e^{\mathrm{j}2\pi{f_{D,2}}\left(t-\hat\tau_{\textrm{ind}_2}\right)}e^{\mathrm{j}{\phi_{n_{I_m}}}},
 \label{eq:indirect_2}
\end{align}\normalsize
where {$\hat\tau_{\textrm{ind}_2}=\tau_{\textrm{ind}_2, n_{I_m}, n_{B_t}, n_{B_r}}$ and} $\phi_{n_{I_m}}$ denotes the phase shift of the $n$-th element of the $m$-th IRS. Then, \textit{mutatis mutandis}, the expressions for the indirect paths $3$ and $4$ are obtained. The overall indirect path signal via $m$-th IRS is a superposition of all three paths, i.e.,
\begin{align}
    s_{\textrm{ind}, I_m, n_{B_r}}(t) = & s_{{\textrm{ind}_2,I_m,n_{B_r}}(t)} \!+ \!s_{{\textrm{ind}_3,I_m,n_{B_r}}(t)} \!+\! s_{{\textrm{ind}_4,I_m,n_{B_r}}(t)}.
\end{align}\normalsize
As in the direct path processing, after downconverting the indirect path receive signal, we similarly combine some terms with the channel gains to obtain new coefficients corresponding to the indirect paths $2$, $3$, and $4$ of the $m$-th IRS as, respectively,
\begin{subequations} \label{channel_gain_ind}
\begin{align} 
 &\widetilde{\alpha}_{I_m,2,k} \!= \!{\alpha}_{I_m,2}e^{\mathrm{j}2\pi{f_{D_2,I_m}}\triangle{t}}e^{-\mathrm{j}2\pi(f_c+f_k)(\tau_{BI_m}+\tau_{TI_m}+\tau_{BT})}, \\
 &\widetilde{\alpha}_{I_m,3,k} \!=\!{\alpha}_{I_m,3}e^{\mathrm{j}2\pi{f_{D_3,I_m}}\triangle{t}}e^{-\mathrm{j}2\pi(f_c+f_k)(\tau_{BI_m}+\tau_{TI_m}+\tau_{BT})}, \\
 &\widetilde{\alpha}_{I_m,4,k} \!=\!{\alpha}_{I_m,4}e^{\mathrm{j}2\pi{f_{D_4,I_m}}\triangle{t}}e^{-\mathrm{j}2\pi(f_c+f_k)(2\tau_{BI_m}+2\tau_{TI_m})}.
 \end{align}
\end{subequations}\normalsize 
 The channel matrices for the paths $m$-th IRS-target-Rx,  $m$-th IRS-Rx, Tx-target-$m$-th IRS, Tx-$m$-th IRS, and $m$-th IRS-target-$m$-th IRS are, respectively,
 \begin{subequations}
\begin{align}%\label{10}
  {\bf E}_{m,k} &= {\bf a}_{B_r}(\theta_{BT},f_k){\bf a}_{I_m}^T(\theta_{I_mT},f_k), \\
 {\bf D}_{m,k}  &= {\bf a}_{B_r}(\theta_{BI_m},f_k){\bf a}_{I_m}^T(\theta_{I_mB},f_k), \\
 {\bf B}_{m,k} &= {\bf a}_{I_m}(\theta_{I_mT},f_k){\bf a}_{B_t}^T(\theta_{BT},f_k), \\
 {\bf G}_{m,k} & = {\bf a}_{I_m}(\theta_{I_mB},f_k){\bf a}_{B_t}^T(\theta_{BI_m},f_k),\\
 {\bf W}_{m,k} &= {\bf a}_{I_m}(\theta_{I_mT},f_k){\bf a}_{I_m}^T(\theta_{I_mT},f_k),
\end{align}
\end{subequations}\normalsize
where $\theta_{I_mB}$, $\theta_{I_mT}$ and $\theta_{BI_m}$ denote the angle of $m$-th IRS with respect to DFBS, target, and DFBS with respect to $m$-th IRS, respectively. Hereafter, in the channel matrices, the subscripts $m$ and $k$ denote the $m$-th IRS and $k$-th subcarrier.

We exploit the similarities between \eqref{eq:indirect_2} and \eqref{3} and long enough cyclic prefix to omit the intermediate steps and present the resulting simplified models in the sequel. 
It follows from \eqref{6} that, in frequency-domain, the received signal at $k$-th subcarrier via direct path is the product of the channel response matrix ${\bf A}_{
\textrm{dir},k}$ and the frequency-dependent signal ${\bf F}_k{\bf s}_k$. Therefore, the $N_{B_r} \times 1$ received signal vector for all  receive antennas via indirect path assisted by the $m$-th IRS in the frequency-domain is

\begin{align}
  { \widetilde{\bf s}}_{\textrm{ind},I_m}[f_k]=&(\widetilde{\alpha}_{I_m,2,k}{\bf E}_{m,k}{\bf\Phi}_m{\bf G}_{m,k} 
  +\widetilde{\alpha}_{I_m,3,k}{\bf D}_{m,k}{\bf\Phi}_m{\bf B}_{m,k} \nonumber \\
  &\!+\!\widetilde{\alpha}_{I_m,4,k}{\bf D}_{m,k}{\bf\Phi}_m{\bf W}_{m,k}{\bf\Phi}_m{\bf G}_{m,k}){\bf F}_k{\bf s}_k,
\end{align}\normalsize
where ${\bf\Phi}_m=\mathrm{diag}(e^{\mathrm{j}\phi_{1}},\cdots,e^{\mathrm{j}\phi_{N_{I_m}}})$ denotes the phase-shift matrix of $m$-th IRS. Using the superposition principle, we obtain the receive signal across all $M$ IRSs as
\begin{align}\label{7}
  { \widetilde{\bf s}}_{\textrm{ind}}[f_k]=\sum_{m=1}^{M}{ \widetilde{\bf s}}_{\textrm{ind},I_m}[f_k] 
 = {\bf A}_{\textrm{ind},k}({\bf\Phi}){\bf F}_k{\bf s}_k, 
\end{align}\normalsize
where ${\bf A}_{\textrm{ind},k}({\bf\Phi}) \!=\!\sum_{m=1}^{M} (\widetilde{\alpha}_{I_m,2,k}{\bf E}_{m,k}{\bf\Phi}_m{\bf G}_{m,k} 
  +\widetilde{\alpha}_{I_m,3,k}{\bf D}_{m,k}{\bf\Phi}_m{\bf B}_{m,k} +\widetilde{\alpha}_{I_m,4,k}{\bf D}_{m,k}{\bf\Phi}_m{\bf W}_{m,k}{\bf\Phi}_m{\bf G}_{m,k})$ denotes the indirect IRS-aided  channel response and ${\bf\Phi}\in\{{\bf\Phi}_1,\cdots,{\bf\Phi}_M\}$. 

Similarly, the response matrices of the clutter scatterers for the paths $m$-IRS--DFBS, DFBS--$m$-IRS, and $m$-IRS--clutters--$m$-IRS, respectively, are $\widetilde{\bf E}_{m,k} \!=\! \sum_{q=1}^{Q}{\bf a}_{B_r}(\theta_{C_qT},f_k){\bf a}_{I_m}^T(\theta_{I_mC_q},f_k)$, 
  $\widetilde{\bf B}_{m,k} \!=\! \sum_{q=1}^{Q}{\bf a}_{I_m}(\theta_{I_mC_q},f_k){\bf a}_{B_t}^T(\theta_{BC_q},f_k),$
and $\widetilde{\bf W}_{m,k} = \sum_{q=1}^{Q}{\bf a}_{I_m}(\theta_{I_mC_q},f_k){\bf a}_{I_m}^T(\theta_{I_mC_q},f_k)$.
Hence, the echo signal of the clutters via direct and indirect paths are 
\begin{subequations} \label{channel_gain_clutter}
\begin{align}
 \widetilde{\bf c}_{\textrm{dir}}[f_k] &\!=\! \sum_{q=1}^{Q}{\alpha}_{C_q,1}{\bf a}_r(\theta_{c_q},f_k){\bf a}_t^T(\theta_{c_q},f_k){\bf F}_k{\bf s}_k \nonumber\\
  &=\widetilde{\bf A}_{\textrm{dir},k}{\bf F}_k{\bf s}_k, \\
 \widetilde{\bf c}_{\textrm{ind}}[f_k] \!=&\! \sum_{m=1}^{M}({\alpha}_{C_q,I_m,2}\widetilde{\bf E}_{m,k}{\bf\Phi}_m{\bf G}_{m,k}\!+\! {\alpha}_{C_q,I_m,3}{\bf D}_{m,k}{\bf\Phi}_m\widetilde{\bf B}_{m,k} 
 \nonumber\\
 &\!+\!{\alpha}_{C_q,I_m,4}{\bf D}_{m,k}{\bf\Phi}_m\widetilde{\bf W}_{m,k}{\bf\Phi}_m{\bf G}_{m,k}){\bf F}_k{\bf s}_k       \nonumber\\
       =&\widetilde{\bf A}_{\textrm{ind},k}({\bf\Phi}) {\bf F}_k{\bf s}_k,
 \end{align}
\end{subequations}\normalsize
{where ${\alpha}_{C_q,1}$, ${\alpha}_{C_q,I_m,2}$, ${\alpha}_{C_q,I_m,3}$ and ${\alpha}_{C_q,I_m,4}$ are the channel gain of clutters,  respectively.}

Notice that from \eqref{dop1}-\eqref{dop4} the Doppler frequency is linearly proportional to the target velocity. We discretize the Doppler grid into $P$ points so that the target velocity at any $p$-th grid point is  ${v}_{p}\in[0,{v}_{\textrm{max}}], p=1,\cdots,P$, where ${v}_{\textrm{max}}$  
is the maximum unambiguous velocity of radar detection. {Following  \eqref{dop1}-\eqref{dop4}, denote the respective Doppler slices or shifts corresponding to discretized velocity grid by $f_{D_1,p}$, $f_{D_2,I_m,p}$, $f_{D_3,I_m,p}$, and $f_{D_4,I_m,p}$ \cite{aubry2015optimizing}.}
The composite received radar signal from both direct and indirect paths at $p$-th Doppler slice is 
\begin{equation}\label{13}
  \widetilde{\bf y}_{R_p}[f_k]=\widetilde{\bf s}_{{\mathrm{dir}}_{p}}[f_k]+\widetilde{\bf s}_{\mathrm{ind}_p}[f_k]
  +\widetilde{\bf c}_{\mathrm{dir}}[f_k]+\widetilde{\bf c}_{\mathrm{ind}}[f_k]+\widetilde{\bf n}_R[f_k],
\end{equation}\normalsize
where $\widetilde{\bf n}_R[f_k]\in\mathbb{C}^{N_{B_r}\times 1}$ denotes the radar noise at the $k$-th subcarrier with zero mean and covariance $\sigma_R^2{\bf I}_{N_{B_r}}$. 
{
\begin{remark}
Similar to the indirect path, it is possible to obtain received signal at the radar due to reflections from multiple IRS. Further, assuming that the cyclic prefix of the OFDM is long enough compared to the delay of the significant reflections, it is possible to include these terms appropriately in the matrix ${\bf A}_{\mathrm{ind},k}(\Phi)$ of equation \eqref{7}. As a result, the current system model can be extended to consider additional reflections with book-keeping.  Finally, the precoder design optimization outlined in the paper holds for the multiple reflections as well.    
\end{remark}}

%{\color{blue} {\bf Remark 1:} \textit{Similar to the indirect path, it is possible to obtain received signal at the radar due to reflections from multiple IRS. Further, assuming that the cyclic prefix of the OFDM is long enough compared to the delay of the significant reflections, it is possible to include these terms appropriately in the matrix ${\bf A}_{\mathrm{ind},k}(\Phi)$ of equation \eqref{7}. As a result, the current system model can be extended to consider additional reflections with book-keeping.  Finally, the precoder design optimization outlined in the paper holds for the multiple reflections as well.}}

{For the radar system, the performance of target detection is largely determined by the output SINR and the detection performance for a given false-alarm improves with the increasing of SINR. Thus, the maximization of SINR is widely used as the optimization criterion \cite{tsinos2021joint}.}
From \eqref{13}, the average SINR of radar for the $p$-th Doppler slice is 
\begin{align}
 &\mathrm{SINR}_{R_p} = \frac{\sum_{k=1}^{K}|{\bf w}^H_p(\widetilde{\bf s}_{\mathrm{dir}_p}[f_k]+\widetilde{\bf s}_{\mathrm{ind}_p}[f_k])|^2}{\sum_{k=1}^{K}|{\bf w}^H_p(\widetilde{\bf c}_{\mathrm{dir}}[f_k]\!+\!\widetilde{\bf c}_{\mathrm{ind}}[f_k])|^2\!+\!K\sigma_R^2{\bf w}_p^H{\bf w}_p} \nonumber\\
 &
 = \frac{\sum_{k=1}^{K}\|{\bf w}_p^H({\bf A}_{\mathrm{dir},k,p}\!+\!{\bf A}_{\mathrm{ind},k,p}({\bf\Phi})){\bf F}_k\|_2^2}{\sum_{k=1}^{K}\|{\bf w}_p^H(\widetilde{\bf A}_{\mathrm{dir},k}\!+\!\widetilde{\bf A}_{\mathrm{ind},k}({\bf\Phi})){\bf F}_k\|_2^2\!+\!K\sigma_R^2{\bf w}_p^H{\bf w}_p}, \label{14}
\end{align}\normalsize
where ${\bf w}_p$, $p=1\cdots,P$ denotes the Doppler filter bank.

\subsection{Communications receiver}
Following the wideband channel model \cite{park2017dynamic}, we denote the CSI from transmitter and $m$-th IRS to $u$-th user  at the $k$-th subcarrier, respectively, by %${\bf H}_{k}\in\mathbb{C}^{U\times{N_t}}$ and ${\bf H}_{m,k}\in\mathbb{C}^{U\times{N_m}}$, respectively, {which are written as

\begin{align}\label{55}
     &{\bf h}_{u,k} \!=\! \sum_{l=1}^{L}\sum_{k=1}^{K}{\alpha_{l}}e^{-\mathrm{j}\frac{2\pi kd}{K}}{\bf a}_{B_t}(\phi_{l},f_k)r(kT_s-\tau_l),
\end{align}  \normalsize
and
\begin{align}
    &{\bf h}_{u,m,k}  \!=\!\sum_{l_m=1}^{L_m}\sum_{k=1}^{K}{\alpha_{l_m}}e^{-\mathrm{j}\frac{2\pi kd}{K}}{\bf a}_{B_t}(\phi_{l_m},f_k)r(kT_s-\tau_{l_m}),
\end{align}\normalsize
where $\alpha_l$ and $\alpha_{l_m}$ are the channel gains; $L$ and $L_m$ are the number of clusters; $\tau_l$ and $\tau_{l_m}$ are the path delays; $\phi_{l}$ and $\phi_{l_m}$ denote the AoDs of all clusters; and $r(t)$ is the raised-cosine function that takes into account the effect of Nyquist filters employed at the transmitter and receiver to avoid the intersymbol interference as, %
%\begin{align}
%\label{eq:rc2}
$r(t)\sum\limits_{k=-\infty}^{+\infty}\delta(t-k K/\triangle T)=\delta(t)$.
%\end{align}\normalsize

{Several well-known methods to estimate the CSI are available in the literature \cite{wei2021channel,he2021channel}. Hence, in this work, assume CSI is known/estimated \textit{a priori}.} The receive signal of the users at the $k$-th subcarrier is \cite{cheng2021hybrid}
\begin{align} \label{15}
  {y}_{C_u}&(t)=  \sum_{k=1}^{K}({\bf h}_{u,k}^T{\bf F}_k{\bf s}_ke^{\mathrm{j} 2\pi{f_k}t} 
 \nonumber \\
 & +\sum_{m=1}^{M}{\bf h}_{u,m,k}^T{\bf\Phi}_m{\bf G}_{m,k}{\bf F}_k
                    {\bf s}_ke^{\mathrm{j}2\pi{f_k}t} +n_{C_u}(t)),
\end{align}\normalsize
where $n_{C_u}(t)$ denotes the $u$-th users' complex white Gaussian noise with zero mean and covariance $\triangle t\sigma_C^2{\bf I}_{U}$. 
Sampling \eqref{15} at the rate $K/\triangle t$ within the symbol duration, we obtain 
\begin{align}
  y_{C_u}&[\widetilde{n}]
  = \sum_{k=1}^{K}( {\bf h}_{u,k}{\bf F}_k{\bf s}_ke^{\mathrm{j}2\pi{f_k}\frac{\widetilde{n}\triangle t}{K}} \nonumber\\
    &
    +\sum_{m=1}^{M}{\bf h}_{u,m,k}{\bf\Phi}_m{\bf G}_{m,k}{\bf F}_k{\bf s}_ke^{\mathrm{j}2\pi{f_k}\frac{\widetilde{n}\triangle t}{K}}+n_{C_u}[\widetilde{n}]),
\end{align}\normalsize
where $\widetilde{n}=1,\cdots,K$ is the discrete-time sample index and synchronized with the radar receiver. Applying $K$-point FFT along the index $\widetilde{n}$, the receive signal of $u$-th user at $k$-th subcarrier is  
\begin{align} \label{16}
     & \widetilde{\bf y}_{C_u}[f_k]\!=\!\underbrace{{\bf h}^T_{u,k}{\bf f}_{k,u}{\bf s}_{k,u}\!+\!\sum_{m=1}^{M}{\bf h}^T_{u,m,k}{\bf\Phi}_m{\bf G}_{m,k}{\bf f}_{k,u}{\bf s}_{k,u}}_{\textrm{desired signal}} 
   \nonumber\\
  &\!+\!\underbrace{\sum_{i\neq u}{\bf h}^T_{u,k}{\bf f}_{k,i}{\bf s}_{k,i}\!+\!\sum_{i\neq u}\sum_{m=1}^{M}{\bf h}^T_{u,m,k}{\bf\Phi}_m{\bf G}_{m,k}{\bf f}_{k,i}{\bf s}_{k,i}}_{\textrm{MUI}}\!+\!\widetilde{n}_{C_u}[f_k],  
\end{align}\normalsize
where ${\bf f}_{k,u}$ denotes the $u$-th column of ${\bf F}_{k}$ and $\widetilde{n}_{C_u}[f_k]$ denotes the additional noise of $u$-th user at the $k$-th subcarrier with zero mean and variance $\sigma_C^2$.

We define ${\bf\Lambda}_u$ as the diagonal matrix with $u$-th diagonal element is one and the others are zero. Following \eqref{16}, the average signal power of the $u$-th user over all $K$ subcarriers is 
\vspace{-0.2cm}
\begin{equation}\label{17}
  {\mathcal{P}_{u}}=\frac{1}{K}\sum_{k=1}^{K}\|({\bf h}^T_{u,k}+\underbrace{\sum_{m=1}^{M}{\bf h}^T_{u,m,k}{\bf\Phi}_m{\bf G}_{m,k})}_{{\mathbf z}_{u,k}({\bf\Phi})}{\bf F}_{k}{\bf\Lambda}_u\|_2^2.
\vspace{-0.2cm}\end{equation}\normalsize
Similarly, defining $\widetilde{\bf\Lambda}_u$ as the diagonal matrix with $u$-th diagonal element is zero and the others are one, the average power of MUI at the $u$-th user is %
%\begin{equation}\label{18}
   ${\mathcal{P}_{\textrm{MUI}}} =\frac{1}{K}\sum_{k=1}^{K}\|({\bf h}^T_{u,k}+\sum_{m=1}^{M}{\bf h}^T_{u,m,k}{\bf\Phi}_m{\bf G}_{m,k}){\bf F}_{k}\widetilde{\bf\Lambda}_u\|_2^2$.
%\end{equation}\normalsize
The average communications SINR of the $u$-th users is  
\begin{align}
  &\mathrm{SINR}_{C_u} \!=\! \frac{\sum_{k=1}^{K}\|({\bf h}^T_{u,k}+{\bf z}_{u,k}({\bf\Phi})){\bf F}_{k}{\bf\Lambda}_u\|_2^2}{\sum_{k=1}^{K}\|({\bf h}^T_{u,k}+{\bf z}_{u,k}({\bf\Phi})){\bf F}_{k}\widetilde{\bf\Lambda}_u\|_2^2+K\sigma^2_{C}} \nonumber\\
  & 
  \!=\! \frac{\sum_{k=1}^{K}\|\mathrm{vec}({\bf F}_k)^H({\bf\Lambda}_u\otimes({\bf h}^T_{u,k}+{\bf z}_{u,k}({\bf\Phi}))^H)\|_2^2}{\sum_{k=1}^{K}\|\mathrm{vec}({\bf F}_k)^H(\widetilde{\bf\Lambda}_u\otimes({\bf h}^T_{u,k}\!+\!{\bf z}_{u,k}({\bf\Phi}))^H)\|_2^2\!+\!K\sigma^2_{C}}, \label{19}
\end{align}\normalsize
where ${\bf z}_{u,k}({\bf\Phi})\!=\!\sum_{m=1}^{M}{\bf h}^T_{u,m,k}{\bf\Phi}_m{\bf G}_{m,k}$ denotes the multi-IRS-aided channel of {the $u$-th user.}

\subsection{Joint design problem}
Our goal is to design the transmit beamformers ${\bf F}_k$, IRS phase-shifts ${\bf\Phi}_m$, and receive filer bank ${\bf w}_p$ that maximize the minimum radar 
SINR over different Doppler slices while guaranteeing the communications SINR of all users. The resulting optimization problem is 
 
\begin{subequations} \label{20}
  \begin{align}
    \mathop{{\mathrm{maximize}}}\limits_{{\bf w}_p,{\bf\Phi}_m,{\bf F}_k} & {\quad} \min_{p}\{\mathrm{SINR}_{R_p}\} \label{20a}\\
    \textrm{subject to}  & {\quad} \mathrm{SINR}_{C_u} \geq \xi, \forall u, \label{20c} \\
    & {\quad} \|{\bf F}_k\|_F^2 \leq \mathcal{P}_k, \forall k, \label{20b}\\
    & {\quad} |{\bf\Phi}_m(i,i)|=1, \forall i,\forall m, \label{20d}
  \end{align} 
\end{subequations}\normalsize 
where $|{\bf\Phi}_m(i,i)|=1$ denotes the unimodular constraint over all elements of IRS phase-shift matrices and $\xi$ is the threshold of communications SINR. The problem in \eqref{20} is highly nonconvex because of the maximin objective function, difference of convex (DC), and constant modulus constraints. Hence, it is difficult to obtain a closed-form solution directly. %In the next section, we develop an AM-based algorithm to tackle this problem. % \eqref{20} in the sequel. 
{
\begin{remark}
Similar to the numerous works on DFRC precoder design, this paper undertakes the joint optimization over each coherence interval of the channel. The generation of the  channel realizations are presented in Section \uppercase\expandafter{\romannumeral4}.   
\end{remark}}

%{\color{blue} {\bf Remark 2:} \textit{Similar to the numerous works on DFRC precoder design, this paper undertakes the joint optimization over each coherence interval of the channel. The generation of the  channel realizations are presented in Section \uppercase\expandafter{\romannumeral4}.}}

\section{Alternating Optimization}
Even without the constraints \eqref{20c} and \eqref{20d}, the optimization problem in \eqref{20} is still nonconvex with respect to ${\bf w}_p$, ${\bf\Phi}_m$ and ${\bf F}_k$ because of the fractional maximin objective function. Thus, the global optimal solution is intractable \cite{aubry2015optimizing}. Meanwhile, all the design variables including the Doppler filter bank, wideband beamformer and IRS phase-shifts are also coupled. Therefore, we resort to the AO framework \cite{bertsekas1997nonlinear}, also known as block coordinate descent, to decouple the design problems of the Doppler filter bank, wideband beamformer, and phase-shifts. Then, the corresponding subproblems are approximately solved in each iteration. This optimization framework also has the flexibility to apply various update order of the blocks updates to this algorithm. More importantly, if each subproblem is solved optimally or suboptimally so that the objective value is improved, then the monotonicity of AO could be guaranteed. {For a lower bounded objective function in a maximization problem,} this monotonicity will ensure the convergence \cite{cheng2021hybrid,aubry2015optimizing}.  

%Whereas it is usually difficult to prove the algorithm convergence in many nonconvex problems, AO has been shown to have theoretical convergence guarantees under some conditions {which conditions?} {put ziyang's paper here and the other works using AO}.

\subsection{Doppler filter bank design}
Denote the signal covariance matrices of clutter and target at the $p$-th Doppler slice, respectively, by %
%\begin{subequations} \label{22}
%  \begin{align}
    ${\bf\Upsilon}_{c} %& 
    = \sum_{k=1}^{K}(\widetilde{\bf A}_{\mathrm{dir},k}\!+\!\widetilde{\bf A}_{\mathrm{ind},k}({\bf\Phi})){\bf F}_k{\bf F}_k^H(\widetilde{\bf A}_{\mathrm{dir},k}\!+\!\widetilde{\bf A}_{\mathrm{ind},k}({\bf\Phi}))^H$ %,\\
    and
    ${\bf\Upsilon}_{p,t} %& 
    = \sum_{k=1}^{K}({\bf A}_{\mathrm{dir},k,p}+{\bf A}_{\mathrm{ind},k,p}({\bf\Phi})){\bf F}_k{\bf F}_k^H({\bf A}_{\mathrm{dir},k,p}+{\bf A}_{\mathrm{ind},k,p}({\bf\Phi}))^H$. 
%  \end{align}
%\end{subequations}\normalsize
Thus, for a given transmit beamforming ${\bf F}_k$ and phase-shift ${\bf\Phi}_m$, the 
subproblem to obtain Doppler filter bank ${\bf w}_p$ is
\begin{equation}\label{21}
   \mathop{{\mathrm{maximize}}}\limits_{{\bf w}_1,\cdots,{\bf w}_P}~\min_{p}\frac{{\bf w}_p^H{\bf\Upsilon}_{p,t}{\bf w}_p}{{\bf w}_p^H{\bf\Upsilon}_{c}{\bf w}_p+K\sigma_R^2{\bf w}_p^H{\bf w}_p}.
\end{equation}\normalsize
Note that the objective function of problem \eqref{21} is separable in terms of the variables ${\bf w}_p,p=1,\cdots,P$. Hence, we obtain an optimal solution for the maximin problem \eqref{21} by solving the following $P$ disjoint problems
\begin{equation}\label{23}
   \mathop{{\mathrm{maximize}}}\limits_{{\bf w}_p}\quad \frac{{\bf w}_p^H{\bf\Upsilon}_{p,t}{\bf w}_p}{{\bf w}_p^H\widetilde{\bf\Upsilon}_{c}{\bf w}_p}, 
\end{equation}\normalsize
where $p=1,\cdots,P$ and $\widetilde{\bf\Upsilon}_{c}={\bf\Upsilon}_{c}+K\sigma_R^2{\bf I}$. Using the Charnes-Cooper transformation \cite{charnes1968programming}, we convert \eqref{23} to an equivalent problem
\begin{equation}\label{24}
\begin{split}
  \mathop{{\mathrm{minimize}}}\limits_{{\bf w}_p}  & \quad{\bf w}_p^H\widetilde{\bf\Upsilon}_{c}{\bf w}_p \\
 \textrm{subject to}  &\quad {\bf w}_p^H{\bf\Upsilon}_{p,t}{\bf w}_p=1.
\end{split}
\end{equation}\normalsize
This problem is a complex-valued homogeneous QCQP, which is nonconvex because of the quadratic constraint. We use the SDR to reformulate \eqref{24} as
\begin{equation}\label{25}
\begin{split}
  \mathop{{\mathrm{minimize}}}\limits_{{\bf W}_p} & \quad \mathrm{Tr}(\widetilde{\bf\Upsilon}_{c}{\bf W}_p)  \\
    \textrm{subject to} & \quad  \mathrm{Tr}({\bf\Upsilon}_{p,t}{\bf W}_p)=1,\; {\bf W}_p\succeq{\bf 0}.
\end{split}
\end{equation}\normalsize
Note that for a complex-valued homogeneous QCQP with three constraints or less, SDR is tight \cite{huang2010rank}. Denote the optimal solution of complex-value SDR \eqref{25} by ${\bf W}_p^{\star}$, there always exists  $\mathrm{rank}({\bf W}^{\star}_p) \leq \sqrt{n_p}$
where $n_p$ denotes the number of constraint \cite{luo2010semidefinite}. This inequality leads to $\mathrm{rank}({\bf W}^{\star}_p)=1$ in problem \eqref{25}. Hence, the optimal solution of problem \eqref{24} is obtained via eigenvalue decomposition (EVD) as ${\bf W}_p^{\star}={\bf w}_p^{\star}{{\bf w}_p^{\star}}^H$. Note that \eqref{24} can also be solved using the Karush–Kuhn–Tucker (KKT) conditions or even successive convex approximation where the constraint is linearized via the first-order Taylor approximation. In our numerical experiments, we empirically observe that the current SDR approach is already efficient.

\subsection{Transmit beamformer design}
%For wideband transmit beamformers, to simplify the expressions, 
Expanding the SINR expressions in \eqref{14} and \eqref{19}, the square terms in the numerators and denominators yield inner products. We substitute the following expressions
\begin{subequations} \label{28}
  \begin{align}
  &{\bf R}_{u,k} = {\bf\Lambda}_u\otimes({\bf h}^T_{u,k}+{\bf z}_{u,k}({\bf\Phi}))^H({\bf h}^T_{u,k}+{\bf z}_{u,k}({\bf\Phi})),\\
  &\widetilde{\bf R}_{u,k} =\widetilde{\bf\Lambda}_u\otimes({\bf h}^T_{u,k}+{\bf z}_{u,k}({\bf\Phi}))^H({\bf h}^T_{u,k}+{\bf z}_{u,k}({\bf\Phi})), \\
  &{\bf\Xi}_{p,c,k} ={\bf I}_U\!\otimes\!{(\widetilde{\bf A}_{\mathrm{dir},k}\!+\!\widetilde{\bf A}_{\mathrm{ind},k}({\bf\Phi}))^H{\bf w}_p{\bf w}^H_p(\widetilde{\bf A}_{\mathrm{dir},k}\!+\!\widetilde{\bf A}_{\mathrm{ind},k}({\bf\Phi}))},\\
  &{\bf\Xi}_{p,t,k} ={\bf I}_U\!\otimes\!{({\bf A}_{\mathrm{dir},k,p}\!+\!{\bf A}_{\mathrm{ind},k,p}({\bf\Phi}))^H{\bf w}_p{\bf w}^H_p({\bf A}_{\mathrm{dir},k,p}+{\bf A}_{\mathrm{ind},k,p}({\bf\Phi}))},
  \end{align}
\end{subequations}\normalsize
%to the expanded SINR for radar and communications in \eqref{14} and \eqref{19}, respectively. 
in both SINRs, where the Kronecker product follows from the matrix identities \cite[eqs. (520)-(524)]{petersen2008matrix}. Now, the optimization problem to design transmit beamformers with fixed ${\bf w}_p$ and ${\bf\Phi}_m$ becomes
\begin{equation}\label{27}
  \begin{split}
    \mathop{{\mathrm{maximize}}}\limits_{{\bf f}_1,\cdots,{\bf f}_K} & {\quad} \min_{p}\frac{\sum_{k=1}^{K}{\bf f}^H_k{\bf\Xi}_{p,t,k}{\bf f}_k}{\sum_{k=1}^{K}{\bf f}^H_k{\bf\Xi}_{p,c,k}{\bf f}_k+K\sigma_R^2{\bf w}_p^H{\bf w}_p} \\
        \textrm{subject to}          & {\quad} \|{\bf f}_k\|_2^2 \leq \mathcal{P}_k, \forall k, \;\frac{\sum_{k=1}^{K}{\bf f}^H_k{\bf R}_{u,k}{\bf f}_k}{\sum_{k=1}^{K}{\bf f}^H_k
        \widetilde{\bf R}_{u,k}{\bf f}_k+K\sigma^2_{C}}\geq \xi, \forall u,
  \end{split}
\end{equation}\normalsize
where ${\bf f}_k=\mathrm{vec}({\bf F}_k), k=1,\cdots,K$. Next, we replace the summations in \eqref{28} by defining the following block diagonal matrices 
\begin{equation}\label{29}
\begin{split}
  &{\bf R}_{u} =\mathrm{blkdiag}({\bf R}_{u,1},\cdots,{\bf R}_{u,K}),\; \widetilde{\bf R}_{u}  =\mathrm{blkdiag}(\widetilde{\bf R}_{u,1},\cdots,\widetilde{\bf R}_{u,K}), \\
 &{\bf\Xi}_{p,t}  =\mathrm{blkdiag}({\bf\Xi}_{p,t,1},\cdots,{\bf\Xi}_{p,t,K}),\; {\bf\Xi}_{p,c}  =\mathrm{blkdiag}({\bf\Xi}_{p,c,1},\cdots,{\bf\Xi}_{p,c,K}),
\end{split}
\end{equation}\normalsize
According to \eqref{29}, the problem in \eqref{27} is equivalent to % reformulated as
\begin{equation}\label{30}
  \begin{split}
    \mathop{{\mathrm{maximize}}}\limits_{{\mathbf f}\in\mathbb{C}^{KN_{B_t}U\times1}} & {\quad} \min_{p}\frac{{\mathbf f}^H{\bf\Xi}_{p,t}{\mathbf f}}{{\mathbf f}^H{\bf\Xi}_{p,c}{\mathbf f}+K\sigma_R^2{\bf w}_p^H{\bf w}_p} \\
       \textrm{subject to}          & {\quad} \|{\bf V}_{k}{\mathbf f}\|_2^2 \leq \mathcal{P}_k, \forall k, \; \frac{{\mathbf f}^H{\bf R}_{u}{\mathbf f}}{{\mathbf f}^H
                                 \widetilde{\bf R}_{u}{\mathbf f}+K\sigma^2_{C}}\geq \xi, \forall u,
  \end{split}
\end{equation}\normalsize
where ${\mathbf f}=[{\bf f}_1^T,\cdots,{\bf f}_K^T]^T$, ${\bf V}_{k}$ denotes the selection matrix to extract $k$-th interval of ${\mathbf f}$, e.g., the vector ${\bf f}_k$. To tackle the maximin problem \eqref{30}, recall the following lemma about the generalized fractional programming (GFP), which states the requirement and method to achieve the optimal solution.
\begin{lemma}\cite{aubry2015optimizing} Consider two sets $\{g_p({\bf x})\}_{p=1}^P$ and $\{f_p({\bf x})\}_{p=1}^P$ of, respectively, convex and non-negative concave functions over a convex set $\mathcal{X}$. Then, the GFP problem
\begin{equation}\label{31}
    \mathop{{\mathrm{maximize}}}\limits_{{\bf x}} \quad \min_{p}~ \frac{f_p({\bf x})}{g_p({\bf x})}\;
     \; \textrm{subject to}   \quad       {\bf x}\in\mathcal{X},
\end{equation}\normalsize
is solvable and the optimal solution can be obtained via generalized Dinkelbach algorithm.
\label{lem:gfp}
\end{lemma}

It follows from Lemma~\ref{lem:gfp} that the generalized Dinkelbach algorithm is unable to solve the problem \eqref{30} because of the convexity of the numerator in the objective function and the second DC constraint. We aim to linearize these expressions through the following Lemma~\ref{lem:lin}, 
\begin{lemma} For the function $f({\bf x})={\bf x}^H{\bf H}{\bf x}$, the following inequality is always satisfied
\begin{equation} \label{32}
  f({\bf x})\geq 2\Re({{\bf x}^{(n)}}^H{\bf H}{\bf x})-f({{\bf x}^{(n)}}),
\end{equation}\normalsize
where ${\bf H}$ is {positive semidefinite (PSD)} matrix, ${{\bf x}^{(n)}}$ denotes the current point (at the $n$-th iteration), and the equality holds if and only if
${\bf x}={{\bf x}^{(n)}}$.
\label{lem:lin}
\end{lemma}
\begin{IEEEproof}
Define a real-valued function
\begin{equation} \label{33}
   g({\bf x}_r)={\bf x}_r^T{\bf H}_r{\bf x}_r,
\end{equation}\normalsize
where
\begin{equation} \label{34}
{\bf x}_r = \left[\begin{matrix} \Re\{{\bf x}\}\\ \Im\{{\bf x}\} \end{matrix}\right], \quad
{\bf H}_r = \left[\begin{matrix}
                      \Re\{{\bf H}\} & -\Im\{{\bf H}\} \\
                      \Im\{{\bf H}\} &  \Re\{{\bf H}\}  \end{matrix}\right].
\end{equation}\normalsize
Observe that $f({\bf x})=g({\bf x}_r)$ and ${\bf H}_r={\bf H}_r^T$. Then,
\begin{equation} \label{35}
  \begin{split}
      g({\bf x}_r) & \geq g({{\bf x}_{r}^{(n)}})+\nabla^{T}g({{\bf x}_{r}^{(n)}})({\bf x}_r-{{\bf x}_{r}^{(n)}}) \\
                   & ={{\bf x}_{r}^{(n)}}^T{\bf H}_r{{\bf x}_{r}^{(n)}}+{{\bf x}_{r}^{(n)}}^T({\bf H}_r+{\bf H}_r^T)({\bf x}_r-{{\bf x}_{r}^{(n)}}) \\
                   & =2{{{\bf x}_{r}^{(n)}}^T}{\bf H}_r{\bf x}_r-{{\bf x}_{r}^{(n)}}^T{\bf H}_r{{\bf x}_{r}^{(n)}} \\
                   & =2\Re({{\bf x}^{(n)}}^H{\bf H}{\bf x})-g({{\bf x}_{r}^{(n)}}) \\
                   & =2\Re({{\bf x}^{(n)}}^H{\bf H}{\bf x})-f({\bf x}^{(n}),
  \end{split}
\end{equation}\normalsize
where ${{\bf x}_{r}^{(n)}}$ is similarly defined as \eqref{34}. From \eqref{33} and \eqref{35}, we conclude that the inequality \eqref{32} always holds. This completes the proof.
\end{IEEEproof}

Following Lemma 2, we reformulate the problem \eqref{30} and relax it by linearizing the numerator and communications SINR constraint as
\begin{equation}\label{36}
  \begin{split}
    \mathop{{\mathrm{maximize}}}\limits_{{\mathbf f}} & {\quad} \min_{p}\frac{2\Re({{\mathbf f}^{(n)}}^H{\bf\Xi}_{p,t}{\mathbf f})-{{\mathbf f}^{(n)}}^H{\bf\Xi}_{p,t}{{\mathbf f}^{(n)}}}{{\mathbf f}^H{\bf\Xi}_{p,c}{\mathbf f}+K\sigma_R^2{\bf w}_p^H{\bf w}_p} \\
     \textrm{subject to}          & {\quad} \|{\bf V}_{k}{\mathbf f}\|_2^2 \leq \mathcal{P}_k, \forall k,\\
                                 & {\quad} \xi{\mathbf f}^H\widetilde{\bf R}_{u}{\mathbf f}-2\Re({{\mathbf f}^{(n)}}^H{\bf R}_{u}
                                 {\mathbf f})\leq const., \forall u
  \end{split}
\end{equation}\normalsize
where $const.=-\xi K\sigma^2_{C}-{{\mathbf f}^{(n)}}^H{\bf R}_{u}{\mathbf f}^{(n)}$ and $n$ denotes the number of outer iteration. Note that the objective value of \eqref{30} is always greater than or equal to \eqref{36}. Hence, solving \eqref{36} results in approximately solving \eqref{30}. Now, problem \eqref{36} satisfies the requirements of GFP as in \eqref{31}. Thus, we can solve problem \eqref{36} by reformulating it as 
\begin{equation}\label{37}
  \begin{split}
    \mathop{{\mathrm{maximize}}}\limits_{{\mathbf f}} & {\quad} \min_{p} 2\Re({{\mathbf f}^{(n)}}^H{\bf\Xi}_{p,t}{\mathbf f})-\lambda_f{{\mathbf f}^H{\bf\Xi}_{p,c}{\mathbf f}} \\
     \textrm{subject to}          & {\quad} \|{\bf V}_{k}{\mathbf f}\|_2^2 \leq \mathcal{P}_k, \forall k,\\
                                 & {\quad} \xi{\mathbf f}^H\widetilde{\bf R}_{u}{\mathbf f}-2\Re({{\mathbf f}^{(n)}}^H{\bf R}_{u}
                                 {\mathbf f})\leq const., \forall u,
  \end{split}
\end{equation}\normalsize
{where $\lambda_f$ denotes the Dinkelbach parameter.} The equivalent reformulation of this problem is by the epigraph form, which is then solved efficiently. Algorithm~\ref{alg:DBS} summarizes the Dinkelbach-based method. % algorithm for solving problem \eqref{36}.
%-------------------------------------------------
\begin{algorithm}[t]
	\caption{Dinkelbach-based algorithm to solve \eqref{36}}
    \label{alg:DBS}
	\begin{algorithmic}[1]
		\Statex \textbf{Input:}  $\zeta_1$, ${\mathbf f}^{(n)}$ $\mathcal{P}_k$, ${\bf R}_{u}$, $\widetilde{\bf R}_{u}$, ${\bf\Xi}_{p,t}$, and ${\bf\Xi}_{p,c}$. \;
        %\Statex 
        \textbf{Output:}
        ${\mathbf f}^{(n)}$

 \State Set $n_1=0$, ${\mathbf f}_{n_1}={\mathbf f}^{(n)}$;
\State $\lambda_{f}^{(n_1)}=\min_{p}\frac{2\Re({{\mathbf f}^{(n)}}^H{\bf\Xi}_{p,t}{\mathbf f})-{{\mathbf f}^{(n)}}^H{\bf\Xi}_{p,t}{\mathbf f}^{(n)}}{{\mathbf f}^H{\bf\Xi}_{p,c}{\mathbf f}+K\sigma_R^2{\bf w}_p^H{\bf w}_p}$
    \Repeat
      \State Find ${\mathbf f}_{n_1}$ by solving problem \eqref{37} using $\mathcal{P}_k$, ${\bf R}_{u}$, $\widetilde{\bf R}_{u}$, ${\bf\Xi}_{p,t}$, and ${\bf\Xi}_{p,c}$;
      \State 
       $F_{\lambda_{n_1}}=\min_{p} 2\Re({{\mathbf f}^{(n)}}^H{\bf\Xi}_{p,t}{\mathbf f}_{n_1})-\lambda_f^{(n_1)}{{\mathbf f}_{n_1}^H{\bf\Xi}^{c}_{p}{\mathbf f}_{n_1}}$;
      \State $n_1\leftarrow n_1+1$;
      \State Update $\lambda_f^{(n_1)}=\min_{p}\frac{2\Re({{\mathbf f}^{(n)}}^H{\bf\Xi}_{p,t}{\mathbf f}_{n_1})-{{\mathbf f}^{(n)}}^H{\bf\Xi}_{p,t}{\mathbf f}^{(n)}}{{{\mathbf f}_{n_1}}^H{\bf\Xi}_{p,c}{\mathbf f}_{n_1}+K\sigma_R^2{\bf w}_p^H{\bf w}_p}$;
    \Until{$F_{\lambda_{n_1}}\leq \zeta_1$ or reach the maximum iteration. };
    \State \Return ${\mathbf f}^{(n)}={\mathbf f}_{n_1}$;
  \end{algorithmic}
  
\end{algorithm}
%-------------------------------------------------

\vspace{-0.3cm}
\subsection{IRS phase-shifts design}
%In this subsection, we combine the C-ADMM and RSD algorithm to solve the subproblem with the phase-shift design.  
With fixed ${\bf w}_p$ and ${\bf F}_k$, the optimization problem to obtain IRS phase-shifts is 
\begin{equation}\label{38}
  \begin{split}
    \mathop{{\mathrm{maximize}}}\limits_{{\bf\Phi}_m} & {\quad} \min_{p}\frac{\sum_{k=1}^{K}\|{\bf w}_p^H({\bf A}_{\mathrm{dir},k,p}\!+\!{\bf A}_{\mathrm{ind},k,p}({\bf\Phi})){\bf F}_k\|_2^2} {\sum_{k=1}^{K}\|{\bf w}_p^H(\widetilde{\bf A}_{\mathrm{dir},k}\!+\!\widetilde{\bf A}_{\mathrm{ind},k}({\bf\Phi})){\bf F}_k\|_2^2\!+\!K\sigma_R^2{\bf w}_p^H{\bf w}_p} \\
                  \textrm{subject to}          & {\quad} \mathrm{SINR}_{C_u} \geq \xi, \forall u,  \; |{\bf\Phi}_m(i,i)|=1, \forall i, \forall m.
  \end{split}
\end{equation}\normalsize
It is very challenging to directly solve this problem because of the fractional quartic objective function, DC, and constant modulus constraints. Define the vector  ${\bm{\phi}}=[{\bm{\phi}}_1^T,\cdots,{\bm{\phi}}_M^T]^T$, where ${\bm{\phi}}_m={{\bf\Phi}_m}{\bf 1}_{N_{I_m}}$. The cascaded communications channel matrices of $u$-th user is 
\begin{subequations} \label{channel_comm}
  \begin{align}
  &q_u=\sum_{k=1}^{K}{\bf h}_{u,k}^T{\bf F}_k{\bf\Lambda}_u{\bf F}_k^H{\bf h}_{u,k}^{\ast}, 
  {\bf q}_u = \sum_{k=1}^{K}{\bf H}_{u,k}^{\ast}{\bf F}_k^{\ast}{\bf\Lambda}_u{\bf F}_k^T{\bf h}_{u,k}, \\ 
  & {\bf Q}_u=\sum_{k=1}^{K}{\bf H}_{u,k}^{\ast}{\bf F}_k^{\ast}{\bf\Lambda}_u{\bf F}_k^T{\bf H}_{u,k},  
  \overline{q}_u=\sum_{k=1}^{K}{\bf h}_{u,k}^T{\bf F}_k\widetilde{\bf\Lambda}_u{\bf F}_k^H{\bf h}_{u,k}^{\ast}, \\ &\overline{\bf q}_u  = \sum_{k=1}^{K}{\bf H}_{u,k}^{\ast}{\bf F}_k^{\ast}\widetilde{\bf\Lambda}_u{\bf F}_k^T{\bf h}_{u,k},
  \overline{\bf Q}_u  = \sum_{k=1}^{K}{\bf H}_{u,k}^{\ast}{\bf F}_k^{\ast}\widetilde{\bf\Lambda}_u{\bf F}_k^T{\bf H}_{u,k}, 
  \end{align} 
\end{subequations}\normalsize
and ${\bf H}_{u,k}=[\mathrm{diag}({\bf h}_{u,1,k}){\bf G}_{1,k},\cdots,\mathrm{diag}({\bf h}_{u,M,k}){\bf G}_{M,k}]^T.$
%${\bf H}_{u,k}\mathrm{diag}({\bf h}_{u,1,k}){\bf G}_{1,k},\cdots, \mathrm{diag}({\bf h}_{u,M,k}){\bf G}_{M,k}]^T$, 
%  $q_u           
%  = \sum_{k=1}^{K}{\bf h}_{u,k}^T{\bf F}_k{\bf\Lambda}_u{\bf F}_k^H{\bf h}_{u,k}^{\ast}$,   
%  $\overline{q}_u$\\ $= \sum_{k=1}^{K}{\bf h}_{u,k}^T{\bf F}_k\widetilde{\bf\Lambda}_u{\bf F}_k^H{\bf h}_{u,k}^{\ast}$, %\nonumber\\
%  ${\bf q}_u     %& 
%  = \sum_{k=1}^{K}{\bf H}_{u,k}^{\ast}{\bf F}_k^{\ast}{\bf\Lambda}_u{\bf F}_k^T{\bf h}_{u,k}$, 
%  $\overline{\bf q}_u  = \sum_{k=1}^{K}{\bf H}_{u,k}^{\ast}{\bf F}_k^{\ast}\widetilde{\bf\Lambda}_u{\bf F}_k^T{\bf h}_{u,k}$, %\nonumber\\
%  ${\bf Q}_u     %& 
%  = \sum_{k=1}^{K}{\bf H}_{u,k}^{\ast}{\bf F}_k^{\ast}{\bf\Lambda}_u{\bf F}_k^T{\bf H}_{u,k}$, and
%  $\overline{\bf Q}_u  = \sum_{k=1}^{K}{\bf H}_{u,k}^{\ast}{\bf F}_k^{\ast}\widetilde{\bf\Lambda}_u{\bf F}_k^T{\bf H}_{u,k}$. %\nonumber
%\end{align}\normalsize
Then, we introduce the auxiliary variables ${\bm\psi}={\bm\phi}$ 
to convert problem \eqref{38} into the bi-fractional quadratic programming as
\begin{align}\label{39}
    \mathop{{\mathrm{maximize}}}\limits_{{\bm\phi},{\bm\psi}} & ~ \min_{p}\frac{\overline{f}_p({\bm\phi},{\bm\psi})}{\overline{g}_p({\bm\phi},{\bm\psi})} \nonumber\\
                  \textrm{subject to}         & ~ {\bm\phi}={\bm\psi},   |{\bm\phi}(i)|=1,|{\bm\psi}(i)|=1,\forall i \nonumber\\
                                 & ~ \frac{q_u+2\Re\{{\bm\phi}^H{\bf q}_{u}\}\!+\!{\bm\phi}^H{\bf Q}_u{\bm\phi}}
                                     {\overline{q}_u\!+\!2\Re\{{\bm\phi}^H\overline{{\bf q}}_{u}\}\!+\!{\bm\phi}^H\overline{\bf Q}_u{\bm\phi}\!+\!K\sigma_C^2}\geq \!\xi, \forall u,
\end{align}\normalsize
Rewrite the objective function as 
\begin{subequations} \label{41}
\begin{align} 
 {\overline{f}_p({\bm\phi},{\bm\psi})} & ={\bm\phi}^H{\bf U}_p{\bm\phi}+2\Re{({\bm\phi}^H{\bf u}_p)}+u_p  \\
                                  & 
                                  ={\bm\psi}^H{\bf V}_p{\bm\psi}+2\Re{({\bm\phi}^H{\bf v}_p)}+v_p, \\  
 {\overline{g}_p({\bm\phi},{\bm\psi})} & ={\bm\phi}^H\widetilde{\bf U}_p{\bm\phi}+2\Re{({\bm\phi}^H\widetilde{\bf u}_p)}+\widetilde{u}_p \\
                                  & 
                                  ={\bm\psi}^H\widetilde{\bf V}_p{\bm\psi}+2\Re{({\bm\phi}^H\widetilde{\bf v}_p)}+\widetilde{v}_p,  
 \end{align}
\end{subequations}\normalsize 
where 
\begin{align}
    &{\bf U}_{k,p}=%\nonumber\\
    \begin{bsmallmatrix}
                  \mathrm{diag}({\bf w}_p^H{\bf E}_{1,k,p}) {\bf G}_{1,k}+\mathrm{diag}({\bf w}_p^H{\bf D}_{1,k}){\bf B}_{1,k,p}+\mathrm{diag}({\bf w}_p^H{\bf D}_{1,k}){\bf W}_{1,k,p}{\bm\psi}_1{\bf G}_{1,k}\\
                  \vdots \\
                  \mathrm{diag}({\bf w}_p^H{\bf E}_{m,k,p}) {\bf G}_{m,k}+\mathrm{diag}({\bf w}_p^H{\bf D}_{m,k}){\bf B}_{m,k,p}+\mathrm{diag}({\bf w}_p^H{\bf D}_{m,k}){\bf W}_{m,k,p}{\bm\psi}_m{\bf G}_{m,k}
                \end{bsmallmatrix},\nonumber\\ 
&{u}_{p}= \sum_{k=1}^{K}{\bf w}^H_p{\bf A}_{p,\mathrm{dir},k}{\bf F}_k{\bf F}^H_k{{\bf A}_{p,\mathrm{dir},k}}^H{\bf w}_p,\; \nonumber\\
&{\bf u}_{p}= \sum_{k=1}^{K}{\bf U}^{\ast}_{k,p}{\bf F}^{\ast}_k{\bf F}^T_k{{\bf A}_{p,\mathrm{dir},k}}^T{\bf w}^{\ast}_p, \nonumber\\
&{\bf U}_{p}= \sum_{k=1}^{K}{\bf U}^{\ast}_{k,p}{\bf F}^{\ast}_k{\bf F}^T_k{\bf U}^T_{k,p},\;  \\
& {\bf V}_{k,p}=\begin{bsmallmatrix}
                  \mathrm{diag}({\bf w}_p^H{\bf D}_{1,k}{\bf\Phi}_1{\bf W}_{1,k,p}){\bf G}_{1,k}\\
                  \vdots \\
                  \mathrm{diag}({\bf w}_p^H{\bf D}_{m,k}{\bf\Phi}_m{\bf W}_{m,k,p}){\bf G}_{m,k}
                \end{bsmallmatrix}, \;\nonumber\\
&
{\bf V}_{p}= \sum_{k=1}^{K}{\bf V}^{\ast}_{k,p}{\bf F}^{\ast}_k{\bf F}^T_k{\bf V}^T_{k,p},\nonumber\\
&\hspace{2em}\left.
+\sum_{m=1}^{M}\left({\bf G}^T_{m,k}{\bf\Phi}^T_m{\bf E}^T_{m,k,p}+{\bf B}^T_{m,k,p}{\bf\Phi}^T_m{\bf D}^T_{m,k}\right)\right){\bf w}_p^{\ast}, \nonumber\\
&{v}_{p} = \nonumber\\
&
\sum_{k=1}^{K}\left\|{\bf w}_p^H{\bf A}_{p,\mathrm{dir},k}{\bf F}_k+\sum_{m=1}^{M}{\bf w}_p^H({\bf E}_{m,k,p}{\bf\Phi}_m{\bf G}_{m,k}+{\bf D}_{m,k}{\bf\Phi}_m{\bf B}_{m,k,p}){\bf F}_k\right\|^2.\nonumber
\end{align}\normalsize
Similarly, $\widetilde{\bf U}_p$, $\widetilde{\bf V}_p$,  $\widetilde{\bf u}_p$, $\widetilde{\bf v}_p$, $\widetilde{u}_p$, and $\widetilde{v}_p$ are also analogously defined by replacing ${\bf A}_{\mathrm{dir},k}$, ${\bf E}_{m,k,p}$, ${\bf B}_{m,k,p}$, and ${\bf W}_{m,k,p}$ with $\widetilde{\bf A}_{\mathrm{dir},k}$, $\widetilde{\bf E}_{m,k,p}$, $\widetilde{\bf B}_{m,k,p}$, and $\widetilde{\bf W}_{m,k,p}$, respectively.
{
\begin{remark}
The phase optimization in \eqref{39} needs to be updated based on the additional multi-IRS reflection terms. Nevertheless, it can be shown that the power received from such multiple reflections is rather small and hence such contributions can be ignored \cite{zhang2021joint}. Particular to the setting in the simulation set up of Section \uppercase\expandafter{\romannumeral4}, the power received for the  Tx-target-IRS1-IRS2-Rx path is 15 dB lower than the Tx-target-IRS1-Rx path.   
\end{remark}}

%{\color{blue} {\bf Remark 3:} \textit{The phase optimization in \eqref{39} needs to be updated based on the additional multi-IRS reflection terms. Nevertheless, it can be shown that the power received from such multiple reflections is rather small and hence such contributions can be ignored \cite{zhang2021joint}. Particular to the setting in the simulation set up of Section \uppercase\expandafter{\romannumeral4}, the power received for the path Tx-target-IRS1-IRS2-Rx is 15 dB lower than the path Tx-target-IRS1-Rx.}}

Following Dinkelbach framework, \eqref{39} becomes
\begin{equation}\label{42}
  \begin{split}
    \mathop{{\mathrm{maximize}}}\limits_{{\bm\phi},{\bm\psi}} & ~ \min_{p} {\overline{f}_p({\bm\phi},{\bm\psi})}-{\lambda}_{\bm\phi}{\overline{g}_p({\bm\phi},{\bm\psi})} \\
                   \textrm{subject to}          & ~ {\bm\phi}={\bm\psi},
                                  |{\bm\phi}(i)|=1,|{\bm\psi}(i)|=1,\forall i,\\
                                 & ~ \frac{q_u\!+\!2\Re\{{\bm\phi}^H{\bf q}_{u}\}\!+\!{\bm\phi}^H{\bf Q}_u{\bm\phi}}
                                     {\overline{q}_u\!+\!2\Re\{{\bm\phi}^H\overline{{\bf q}}_{u}\}\!+\!{\bm\phi}^H\overline{\bf Q}_u{\bm\phi}\!+\!K\sigma_C^2}\geq \!\xi, \!\forall u.
  \end{split}
\end{equation}\normalsize
where ${\lambda}_{\bm\phi}$ denotes the corresponding Dinkelbach parameter. Convert \eqref{42} to the equivalent
\begin{equation}\label{43}
  \begin{split}
    \mathop{{\mathrm{minimize}}}\limits_{{\bm\phi},{\bm\psi}} & ~ \max_{\mathbf\omega} \sum_{p=1}^{P}\omega_p({\lambda}_{\bm\phi}{\overline{g}_p({\bm\phi},{\bm\psi})}-{\overline{f}_p({\bm\phi},{\bm\psi})}) \\
                  \textrm{subject to}          & ~ {\bm\phi}={\bm\psi},|{\bm\phi}(i)|=1,|{\bm\psi}(i)|=1,\forall i\\
                                 & ~ \frac{q_u\!+\!2\Re\{{\bm\phi}^H{\bf q}_{u}\}\!+\!{\bm\phi}^H{\bf Q}_u{\bm\phi}}
                                     {\overline{q}_u\!+\!2\Re\{{\bm\phi}^H\overline{{\bf q}}_{u}\}\!+\!{\bm\phi}^H\overline{\bf Q}_u{\bm\phi}\!+\!K\sigma_C^2}\geq \!\xi, \!\forall u,
  \end{split}
\end{equation}\normalsize
where ${\mathbf\omega}\succeq{\bf0}$ and $\|{\mathbf\omega}\|_1=1$. 

From \cite{yang2020dual} and Lemma~\ref{lem:lin}, we further linearize the communications SINR in \eqref{30} as
\begin{equation}\label{44}
 2\Re\{{\bf r}_u^H{\bm\phi}\} \leq d_u, \forall u,
\end{equation}\normalsize
where ${\bf r}_u=(\xi{\bm\phi}_t^H(\overline{\bf Q}_u-\eta_u{\bf I})+\xi\overline{\bf q}_u^H-{\bf q}_u^H-{\bm\phi}_t^H{\bf Q}_u)^H$, $\eta_u$ is the largest eigenvalue of $\overline{\bf Q}_u$, $d_u=\xi(\overline{q}_u+K\sigma_C^2-2\eta{MN_m}+{\bm\phi}_t^H\overline{\bf Q}_u{\bm\phi}_t)+{\bm\phi}_t^H{\bf Q}_u{\bm\phi}_t-{q}_u$, and ${\bm\phi}_t$ denotes the value of ${\bm\phi}$ at $t$-th iteration. If the inequality \eqref{44} holds, the original SINR inequality constraint is always satisfied. The augmented Lagrangian function of \eqref{43} is 
\begin{equation}\label{45}
  \mathcal{L}({\bm\phi},{\bm\psi},{\bf u},{\bf w},{\rho})=
     f({\bm\phi},{\bm\psi},{\lambda}_{\bm\phi})+\frac{{\rho}}{2}\|{\bm\phi}-{\bm\psi}+{\bf u}\|_2^2 +\Re\{{\bf w}^T{\bf c}\}
\end{equation}\normalsize
where $f({\bm\phi},{\bm\psi},{\lambda}_{\bm\phi})=\max_{\mathbf\omega} \sum_{p=1}^{P}\omega_p({\lambda}_{\bm\phi}{\overline{g}_p({\bm\phi},{\bm\psi})}-{\overline{f}_p({\bm\phi},{\bm\psi})})$, 
$\rho$ is the penalty parameter, ${\bf u}$ and ${\bf w}\succeq{\bf 0}$ denote the auxiliary variables, ${\bf c}=[c_1,\cdots,c_U]^T$ and $c_u=  2\Re\{{\bf r}_u^H{\bm\phi}\} - d_u$. Based on above, 
we summarize our C-ADMM algorithm for solving \eqref{43} 
in  Algorithm~\ref{alg:C-ADMM}.
%-------------------------------------------------
\begin{algorithm}[t]
  \caption{C-ADMM algorithm to solve \eqref{43}}
  \label{alg:C-ADMM}
  \begin{algorithmic}[1]
    \Statex {\bf Input:} $\zeta_2$, ${\bf u}$, ${\bf w}$, ${{\bm\psi}^{(n)}}$ and ${{\bm\phi}^{(n)}}$ \;\;
    %\Statex 
    {\bf Output:} ${{\bm\phi}^{(n)}}^{\star}={\bm\phi}_{n_2}$.
   \State Set $n_2=0$;
    \Repeat
      \State Compute: ${\lambda}_{\bm\phi}^{(n_2)}=\min_{p}\frac{\overline{f}_p({\bm\phi}^{(n)},{\bm\psi}^{(n)})}{\overline{g}_p({\bm\phi}^{(n)},{\bm\psi}^{(n)})}$;
      \State Update ${\bm\phi}_{n_2}$ via solving 
         \begin{equation}\label{46}
           \mathop{{\mathrm{minimize}}}\limits_{{\bm\phi}^{(n)}}~ \mathcal{L}({\bm\phi}^{(n)},{\bm\psi}^{(n)},{\bf u},{\bf w},{\rho}) \quad 
               \textrm{s.t.} ~ |{\bm\phi}^{(n)}|=1.
      \end{equation}\normalsize
      \State Update ${\bm\psi}_{n_2}$ via solving 
      \begin{equation}\label{47}
         \mathop{{\mathrm{minimize}}}\limits_{{\bm\psi}^{(n)}}  ~ \mathcal{L}({\bm\phi}^{(n)},{\bm\psi}^{(n)},{\bf u},{\bf w},{\rho}) \quad 
       \textrm{s.t.}     ~ |{\bm\psi}^{(n)}|=1.
      \end{equation}\normalsize
      \State Update the dual variable ${\bf u}$ and ${\bf w}$;
     \State $n_2\leftarrow n_2+1$;
    \Until{$\|{\bm\phi}_{n_2}-{\bm\phi}_{n_2-1}\|_2^2\leq\zeta_2$  or reach the maximum iteration};
    \State \Return ${\bm\phi}_{n_2}$;
  \end{algorithmic}
\end{algorithm}
%-------------------------------------------------

Based on above, the subproblem in Step 4 of Algorithm~\ref{alg:C-ADMM} %\eqref{46} 
becomes
\begin{equation}\label{48}
  \begin{split}
    \mathop{{\mathrm{minimize}}}\limits_{{\bm\phi}} & ~ \max_{\mathbf\omega} \sum_{p=1}^{P}\omega_p({\lambda}_{\bm\phi}{\overline{g}_p({\bm\phi},{\bm\psi})}-{\overline{f}_p({\bm\phi},{\bm\psi})}) \\
     &\quad\quad
     +\frac{{\rho}}{2}\|{\bm\phi}-{\bm\psi}+{\bf u}\|_2^2 +\Re\{{\bf w}^T{\bf c}\} \\
    \textrm{subject to}          & ~ |{\bm\phi}|=1.
  \end{split}
\end{equation}\normalsize
It is still difficult to apply the RSD algorithm to solve problem \eqref{48} because of the concave function ${\overline{f}_p({\bm\phi},{\bm\psi})}$ in the objective. We utilize Lemma~\ref{lem:lin} to linearize it as
\begin{equation}\label{49}
{\overline{f}_p({\bm\phi},{\bm\psi})}\geq2\Re\{{\bm\phi}^H({\bf U}_p{\bm\phi}^{(n)}+{\bf u}_p)\}+{{\bm\phi}^{(n)}}^H{\bf U}_p{\bm\phi}^{(n)}+u_p.
\end{equation}\normalsize
{Substituting ${\overline{f}_p({\bm\phi},{\bm\psi})}$ by its lower bound in \eqref{49}, problem \eqref{48} becomes}
\begin{equation}\label{50}
  \begin{split}
    \mathop{{\mathrm{minimize}}}\limits_{{\bm\phi}} & ~ \max_{\mathbf\omega} \sum_{p=1}^{P}\omega_p({\lambda}_{\bm\phi}{\overline{g}_p({\bm\phi},{\bm\psi})}\!-\!2\Re\{{\bm\phi}^H({\bf U}_p{\bm\phi}^{(n)}\!+\!{\bf u}_p)\})\\
     &\quad\quad
     +\frac{{\rho}}{2}\|{\bm\phi}-{\bm\psi}+{\bf u}\|_2^2 +\Re\{{\bf w}^T{\bf c}\} \\
    \textrm{subject to}          & ~ |{\bm\phi}|=1.
  \end{split}
\end{equation}\normalsize

%-------------------------------------------------
\begin{algorithm}[t]
    \caption{RSD algorithm for manifold optimization}
    \label{alg:RSD}
	\begin{algorithmic}[1]
		\Statex \textbf{Input:} $\lambda_{\phi}^{(n_2)}$, ${\bf u}$, ${\bf w}$, ${{\bm\psi}^{(n)}}$ and ${{\bm\phi}^{(n)}}$. \;\;
        %\Statex 
        \textbf{Output:} ${\bm\phi}_{n_2}={\bm\phi}_{n_3}$.
    \State Set $n_3=0$, ${\bm\phi}_{n_3}={\bm\phi}_{n}$;
    \Repeat
      \State Calculate ${\mathbf\omega}$ by solving \eqref{50} with fixed ${\bm\phi}_{n_3}$;
      \State Linearize the function  ${\overline{f}_p({\bm\phi},{\bm\psi})}$ using \eqref{49} and ${{\bm\phi}^{(n)}}$;
      \State Compute Euclidean gradient $\nabla{\mathcal{L}({\bm\phi}_{n_3})}$ as \eqref{51} using ${\bf u}$, ${\bf w}$ and ${{\bm\psi}^{(n)}}$;
      \State Compute Riemannian gradient as $\mathrm{grad}{\mathcal{L}({{\bm\phi}_{n_3}})}$ as \eqref{52};
      \State Update ${\bm\phi}_{n_3}$ via the retraction procedure as \eqref{53};
      \State $\lambda_{\phi}^{(n_3)}\leftarrow\min_{p}\frac{\overline{f}_p({\bm\phi},{\bm\psi})}{\overline{g}_p({\bm\phi},{\bm\psi})}$;      
      \State $n_3\leftarrow n_3+1$;
    \Until{$\lambda_{\phi}^{(n_3)}\geq\lambda_{\phi}^{(n_2)}$ or maximum iteration reached};
    \State \Return ${\bm\phi}_{n_3}$;
  \end{algorithmic}
\end{algorithm}
%-------------------------------------------------

It follows that problem \eqref{50} is the manifold optimization problem, which is solved by the RSD algorithm \cite{wang2021joint}. {In order to obtain the Riemannian gradient, we first calculate the Euclidean gradient of the objective function in \eqref{50} as} 
\begin{equation}\label{51}
  \begin{split}
{\nabla_E}{\mathcal{L}({\bm\phi})}=& {\lambda}_{\bm\phi}(2\widetilde{\bf U}_p{\bm\phi}+2\widetilde{\bf u}_p)-2({\bf U}_p{\bm\phi}^{(n)}+{\bf U}_p)\\
&+\rho({\bm\phi}-{\bm\psi}-{\bf u})
+\sum_{u=1}^{U}w_u({2\overline{\bf Q}_u{\bm\phi}+2\overline{\bf q}_u}).
 \end{split}
\end{equation}\normalsize
Then, the Riemannian gradient is obtained by projecting ${\nabla_E}{\mathcal{L}({\bm\phi})}$ into the tangent space as 
\begin{equation}\label{52}
{\nabla_R}{\mathcal{L}({{\bm\phi}})}= {\nabla_E}{\mathcal{L}({\bm\phi})}-\Re\{{\nabla_E}{\mathcal{L}({\bm\phi})}\odot{\bm\phi}\}\odot{\bm\phi}.
\end{equation}\normalsize
Now, we update it by retracting ${\bm\phi}$ into the complex circle manifold as
\begin{equation}\label{53}
{\bm\phi}=\frac{{\bm\phi}-\alpha{\nabla_R}{\mathcal{L}({\bm\phi})}}{|{\bm\phi}-\alpha{\nabla_R}{\mathcal{L}({\bm\phi})}|}.
\end{equation}\normalsize
In each C-ADMM iteration, the subproblem related with ${\bm\phi}$ is solved by Algorithm 3.

Similarly, the subproblem in Step 4 of Algorithm~\ref{alg:C-ADMM} %\eqref{46} 
is 
\begin{equation}\label{54}
  \begin{split}
    \mathop{{\mathrm{minimize}}}\limits_{{\bm\psi}} & ~ \max_{\mathbf\omega} \sum_{p=1}^{P}\omega_p({\lambda}_{\bm\phi}{\overline{g}_p({\bm\phi},{\bm\psi})}-{\overline{f}_p({\bm\phi},{\bm\psi})})\\
     &\quad\quad
     +\frac{{\rho}}{2}\|{\bm\phi}-{\bm\psi}+{\bf u}\|_2^2 \\
    \textrm{subject to}          & ~ |{\bm\psi}|=1,
  \end{split}
\end{equation}\normalsize
which is also solved via Algorithm~\ref{alg:RSD} with the change of the optimization variable.

To summarize, we utilize SDR  {with the {closed-form} solution} for the receive filter bank design. Then, the Dinkelbach-based method yields the transmit beamformers. Finally, the C-ADMM updates the phase-shifts ${\bf\Phi}_m$. Based on above discussion, the overall AM-based procedure is summarized in Algorithm~\ref{alg:AM}.

%-------------------------------------------------
\begin{algorithm}[t]
  \caption{Alternating maximization algorithm to solve \eqref{20}}
  \label{alg:AM}
  \begin{algorithmic}[1]
		\Statex \textbf{Input:} $\zeta_3$, ${\bf w}_p^{(n)}$, ${\mathbf f}^{(n)}$, and ${\bm\phi}^{(n)}$. \;\;
		%\Statex 
		\textbf{Output:} ${\bf w}_p^{\star}={\bf w}^{(n)}_p$, ${\bf F}_k^{\star}={\bf F}_k^{(n)}$ and ${\bm\phi}_m^{\star}={\bm\phi}_m^{(n)}$. 
    \State Set $n=0$;
    \Repeat
      \State Update ${\bf W}^{(n)}_p, p=1,\cdots,P$ via solving problem \eqref{25};
      \State Update ${\bf w}^{(n)}_p$ as the principle eigenvector of ${\bf W}^{(n)}_p$;
      \State Update ${\mathbf f}^{(n)}$ via \textbf{Algorithm~\ref{alg:DBS}}, and
       reconstruct ${\bf F}_k^{(n)}, k=1,\cdots,K$ via ${\mathbf f}^{(n)}$;
      \State Update ${\bm\phi}^{(n)}$ via \textbf{Algorithms~\ref{alg:C-ADMM}} and \textbf{\ref{alg:RSD}};%, and problem \eqref{46} and \eqref{47} are solved by \textbf{Algorithm 3};
      \State Reconstruct ${\bm\phi}_m^{(n)}, m=1,\cdots,M$ via ${\bm\phi}^{(n)}$;      
      \State $n\leftarrow n+1$;
    \Until{$\|{\mathbf f}^{(n)}-{\mathbf f}^{(n-1)}\|_2^2\leq\zeta_3$ or maximum iterations reached}; 
    \State \Return ${\bf w}^{(n)}_p$, ${\bf F}_k^{(n)}$, and ${\bm\phi}_m^{(n)}$;
  \end{algorithmic}
\end{algorithm}
%-------------------------------------------------

\vspace{-0.3cm}
\subsection{Computational complexity}
{The overall computational burden of Algorithm 4 is linear with the number of outer iterations. Meanwhile, at each outer iteration, the closed-form solution of Doppler filter ${\bf w}_p, p\!=\!1,\cdots,P$ is given by solving problem \eqref{25} with the complexity of $\mathcal{O}(PN_{B_r}^{3.5})$. Then, to update the transmit beamforming matrix ${\bf F}_k, k\!=\!1,\cdots,K$, the computational cost of Algorithm 1 is linear with the number of inner iterations $N_1$. At each inner iteration of the Dinkelbach-based method, problem \eqref{37} is solved by the CVX \cite{grant2009cvx} with the complexity of $\mathcal{O}(K^3N_{B_t}^{3}U^{3})$. In order to update the phase-shift matrix ${\bf\Phi}_m, m\!=\!1,\cdots,M$, the C-ADMM and RSD algorithm are combined with the total complexity $\mathcal{O}(N_2(2N_3M^2N_{I_m}^2+M^2N_{I_m}^2))$, where $N_2$ and $N_3$ denote the maximum iteration number of C-ADMM and RSD, respectively. Finally, the total complexity of the proposed algorithm is $\mathcal{O}(PN_{B_r}^{3.5}+N_1K^3N_{B_t}^{3}U^{3}+N_2(2N_3M^2N_{I_m}^2+M^2N_{I_m}^2))$ for each outer iteration.}

\section{Numerical Experiments}
{{\bf System Settings:} Unless otherwise specified, throughout all experiments, the dual-function transmitter and radar receiver are equipped with a {uniform linear array (ULA)} comprising $N_{B_t}=5$ and $N_{B_r}=5$ elements, respectively. The location of DFBS is set to ${\bf p}_B=[0,0]$. A single fast moving target is located at ${\bf p}_T=[20~\mathrm{m},20~\mathrm{m}]$ with the speed ${\bf v}=[v_x,v_y]$, where the intervals $v_x\in(10~\mathrm{m/s},50~\mathrm{m/s}]$ and $v_y\in(20~\mathrm{m/s},60~\mathrm{m/s}]$ are uniformly divided into $P=5$ discrete grid points. Two stationary clutter scatterers are located at 
${\bf p}_{C_1}=[40~\mathrm{m},38~\mathrm{m}]$ and ${\bf p}_{C_2}=[100~\mathrm{m},30~\mathrm{m}]$, respectively. Three IRSs located at ${\bf p}_{I_1}=[48\mathrm{m},40\mathrm{m}]$, ${\bf p}_{I_2}=[60\mathrm{m},40\mathrm{m}]$ and ${\bf p}_{I_3}=[80\mathrm{m},40\mathrm{m}]$ are, respectively, deployed to assist the DFRC system, in which each IRS consists of $50$ reflecting elements. Meanwhile, the dual-function transmitter serves $U=3$ downlink single-antenna users.}

The central frequency of the wideband DFRC is $f_c=10$ GHz and the frequency step of OFDM is set to $\triangle{f}=20$ MHz. The total number of subcarriers is $K=32$. The inter-element spacing for the transmitter, receiver, and IRS arrays, {i.e., $d_{B_t}$, $d_{B_r}$ and $d_{I_m}$,} is set to the half wavelength of the highest frequency, i.e., $d_{B_t,B_r,I_m}={c}/{2f_{max}}$ and $f_{max}=10.32$ GHz to reduce the grating lobes \cite{cheng2021hybrid}. The transmit power at all subcarriers is set to $\mathcal{P}_1=\cdots,=\mathcal{P}_K=1$ dBW. {The noise variances are $\sigma_R^2=-50$ dBm and $\sigma_C^2=-45$ dBm. We set the SINR threshold for all users to $\xi=12$ dB.}
The {initial value of} Doppler receive filter ${\bf w}_p, p=1\cdots,P$ and the transmit beamformers ${\bf F}_k, k=1,\cdots,K$ are randomly generated column vectors and matrices, whose entries follow zero-mean Gaussian distribution and satisfy $\|{\bf F}_k\|_F^2\leq \mathcal{P}_k$. The phase-shift matrices ${\bf\Phi}_m, m=1\cdots,M$ are initialized with the diagonal entries generated from a complex value with unimodulus amplitude and random phase, i.e., $e^{\mathrm{j}\phi_m}$, where $\phi_m\in(0,2\pi]$. 

\noindent\textbf{Communications settings:} For the communications CSI ${\bf h}_{u,k}$ and ${\bf h}_{u,m,k}$ with $u=1,\cdots,U$, we assume $\alpha_l$ and $\alpha_{l_m}\thicksim\mathcal{CN}(0,1)$, the number of clusters are set to $L=15$ and $L_m=15,\forall m$, the pulse shaping function $r(t)$ is modeled as the raised-cosine filter \cite{duggal2020doppler}, the paths delay $\tau_l$ and $\tau_{l_m}$ is uniformly distributed in $[0,KT_s]$ with $T_s=1$, the AoDs of all clusters, i.e., $\phi_{l}$ and $\phi_{l_m}$, are randomly distributed in $(0,2\pi]$. %For single-antenna communications users, the receive steering vector of each user has only one element equal to unity. We stack the channel vector $\mathbf{h}_{u,k}$ for all users into a matrix $\mathbf{H}_k$. This leads to ${\bf a}(\theta,f_k)=[1,\cdots,1]^T\in\mathbb{C}^{U\times1}$.

\noindent\textbf{Radar settings:} We calculate the radar detection probability and false alarm probability based on Monte Carlo simulations.
Specifically, to evaluate the performance of radar detection,  
we first generate the receive signal $\widetilde{\bf y}_{R_p}[f_k]$ in \eqref{13} with the transmit symbol $\mathbb{E}\{{\bf s}_k{\bf s}_k^H\}={\bf I}$ and noise $\widetilde{\bf n}_R[f_k]\in\mathcal{CN}(0,{\sigma}_R^2{\bf I})$. Then, we transfer \eqref{13} to discrete-time domain using $K$-point IFFT and consider the following binary hypothesis problem 
\begin{equation} \label{56}
 \left\{ 
 \begin{aligned}
   \mathcal{H}_0\!:  & ~{\bf y}_{R_0}[\widetilde{n}]={{\bf c}_{\textrm{dir}}[\widetilde{n}]}\!+\!{\bf c}_{\textrm{ind}}[\widetilde{n}]\!+\!{\bf n}_R[\widetilde{n}], \\
   \mathcal{H}_1\!:  & ~{\bf y}_{R_p}[\widetilde{n}]={\bf s}_{\textrm{dir}_p}[\widetilde{n}]\!+\!{\bf s}_{\textrm{ind}_p}[\widetilde{n}] 
  \!+\!{\bf c}_{\textrm{dir}}[\widetilde{n}]\!+\!{\bf c}_{\textrm{ind}}[\widetilde{n}]\!+\!{\bf n}_R[\widetilde{n}],
\end{aligned}
  \right.
\end{equation}\normalsize
where $\widetilde{n}=1,\cdots,K$.
We compute the empirical probabilities of detection $P_{D}$ and false alarm $P_{FA}$ in the post-Doppler-bank stage as follows. \\%and then perform the simple ratio test via comparing the received signal power with the constant threshold $\gamma$. 
\emph{Case 1:}  Under hypothesis $\mathcal{H}_0$, the average power over all subcarriers is
\begin{equation} \label{57}
 \max_{p} ~\sum_{\widetilde{n}=1}^{K}|{\bf w}^H_p{\bf y}_{R_0}[\widetilde{n}]|^2=\widetilde{\beta}_0,
\end{equation}\normalsize
and, for a certain threshold $\gamma$,
\begin{equation} \label{58}
 \left\{ 
 \begin{aligned}
   \mathrm{if}~\widetilde{\beta}_0 > \gamma,  & \quad \mathrm{false~alarm}, \\
   \mathrm{if}~\widetilde{\beta}_0 \leq \gamma,  & \quad \mathrm{no~false~alarm}.
\end{aligned}
  \right.
\end{equation}\normalsize
Based on the above, we count the total number of false alarms $N_{FA}$ to obtain 
%\begin{equation} %\label{59}
$P_{FA}=\frac{N_{FA}}{N_{mont}}$, where $N_{mont}$ denotes the total number of Monte Carlo simulations.
%\end{equation}

\emph{Case 2:}  Under hypothesis $\mathcal{H}_1$, we compute the average power over all subcarriers as 
\begin{equation} \label{60}
 \max_{p} ~\sum_{\widetilde{n}=1}^{K}|{\bf w}^H_p{\bf y}_{R_p}[\widetilde{n}]|^2=\widetilde{\beta}_1,
\end{equation}\normalsize
and 
\begin{equation} \label{61}
 \left\{ 
 \begin{aligned}
   \mathrm{if}~\widetilde{\beta}_1 > \gamma,  & \quad \mathrm{detection}, \\
   \mathrm{if}~\widetilde{\beta}_1 \leq \gamma,  & \quad \mathrm{no~detection}.
\end{aligned}
  \right.
\end{equation}\normalsize
Based on above, we count the total number of detection $N_{{D}}$ to obtain 
$P_{{D}}=\frac{N_{{D}}}{N_{{mont}}}$. Throughout the simulation, we set $N_{mont}=10^{3}$.

{We consider a distance-dependent path loss model with $\alpha = \sqrt{\alpha_{T,C_q}K_0(\frac{r_0}{r})^{\epsilon}}$ for both target and clutter, where $r$ denotes the relative distance of the corresponding path. The signal attenuation is set to $K_0=-30\mathrm{dB}$ at the reference range of $r_0=1\,\mathrm{m}$. The corresponding path loss exponents for target are $\epsilon_{1}=3.2$, $\epsilon_{I_m,2}=3.0$, $\epsilon_{I_m,3}=3.0$ and $\epsilon_{I_m,4}=3.5$, $\forall m$, respectively, for the \textbf{paths 1}, \textbf{2}, \textbf{3}, and \textbf{4}. Meanwhile, the corresponding path loss exponents for clutter are $\epsilon_{C_q,1}=2.4$, $\epsilon_{C_q,I_m,2}=3.0$, $\epsilon_{C_q,I_m,3}=3.0$ and $\epsilon_{C_q,I_m,4}=3.5$, $\forall q$, $\forall m$, respectively, for the \textbf{paths 1}, \textbf{2}, \textbf{3}, and \textbf{4}. Without loss of generality, we set the RCS of targets and clutters as ${\alpha_T}=1$ and ${\alpha_{C_q}}=1$, $\forall q$, respectively.
The {descent step-size} for RSD algorithm is $\alpha=10^{-2.5}$. Finally, we stop the Algorithm~\ref{alg:DBS}, \ref{alg:C-ADMM} and \ref{alg:AM} when the objective value difference of two adjacent iterations $\zeta_1, \zeta_2$ and $\zeta_3\leq10^{-4}$, respectively,
or reach the maximum iteration number $N_1\!=\!20$, $N_2\!=\!20$, $N_3\!=\!150$ and $N=30$ for Algorithm~\ref{alg:DBS}, \ref{alg:C-ADMM}, \ref{alg:RSD} and \ref{alg:AM}, respectively. } 

\subsection{Convergence performance}

\begin{figure}[!t]
    \centering
    \includegraphics[width=0.45\textwidth]{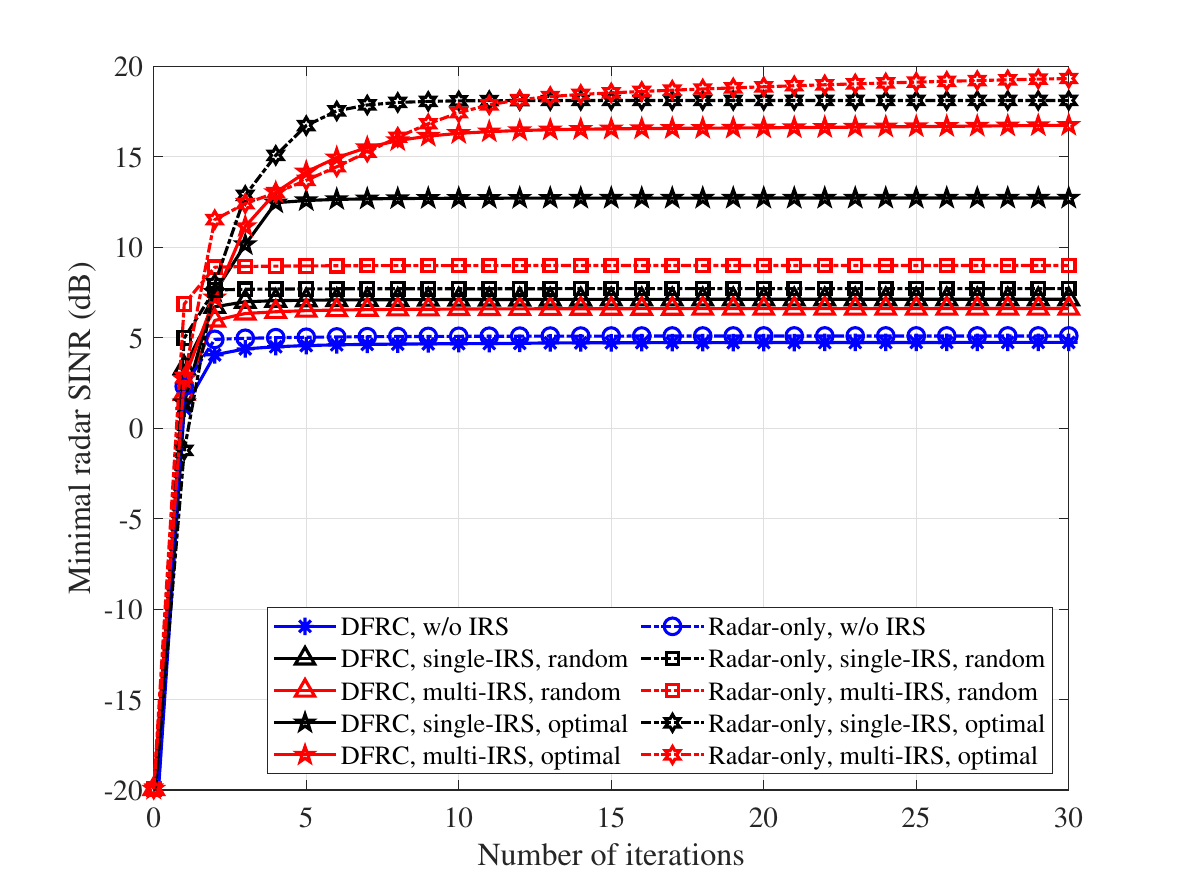}
\caption{Performance of convergence, minimum radar SINR versus the number of iterations.}
\label{fig:convergence}
\end{figure} 
{Fig.~\ref{fig:convergence} demonstrates the convergence performance of proposed AM algorithm. 
For the radar-only system, we remove the communications constraint \eqref{20c} in problem \eqref{20} and then evaluate the minimum radar SINR. We observe that the proposed algorithm converges within 15 outer iterations. The radar-only system without IRS (denoted by “Radar-only, w/o IRS”) has a better performance than the DFRC system without IRS (denoted by “DFRC, w/o IRS”) because the latter needs to allocate a portion of transmit power to serve the users while the radar-only system utilizes all available power for target detection. If we use random phase-shifts for IRS (denoted by “single-IRS, random” or “multi-IRS, random”), the radar SINR is lower than the one obtained after optimizing the phase-shifts (denoted by “single-IRS, optimal” or “multi-IRS, optimal”) but higher than the non-IRS case. Meanwhile, as for DFRC system, optimizing the phase-shifts leads to at least $12$ dB radar SINR gain for multi-IRS deployment compared with the non-IRS case. Finally, the achievable minimum radar SINR for multi-IRS DFRC is higher than “DFRC, w/o IRS”, “radar-only, w/o IRS”  and single-IRS DFRC system because of more reflecting elements and RCS diversity resulting from multi-IRS deployment. }

\subsection{Radar SINR performance}

\begin{figure*}[!t]\centering
\centering
\includegraphics[width=1\textwidth]{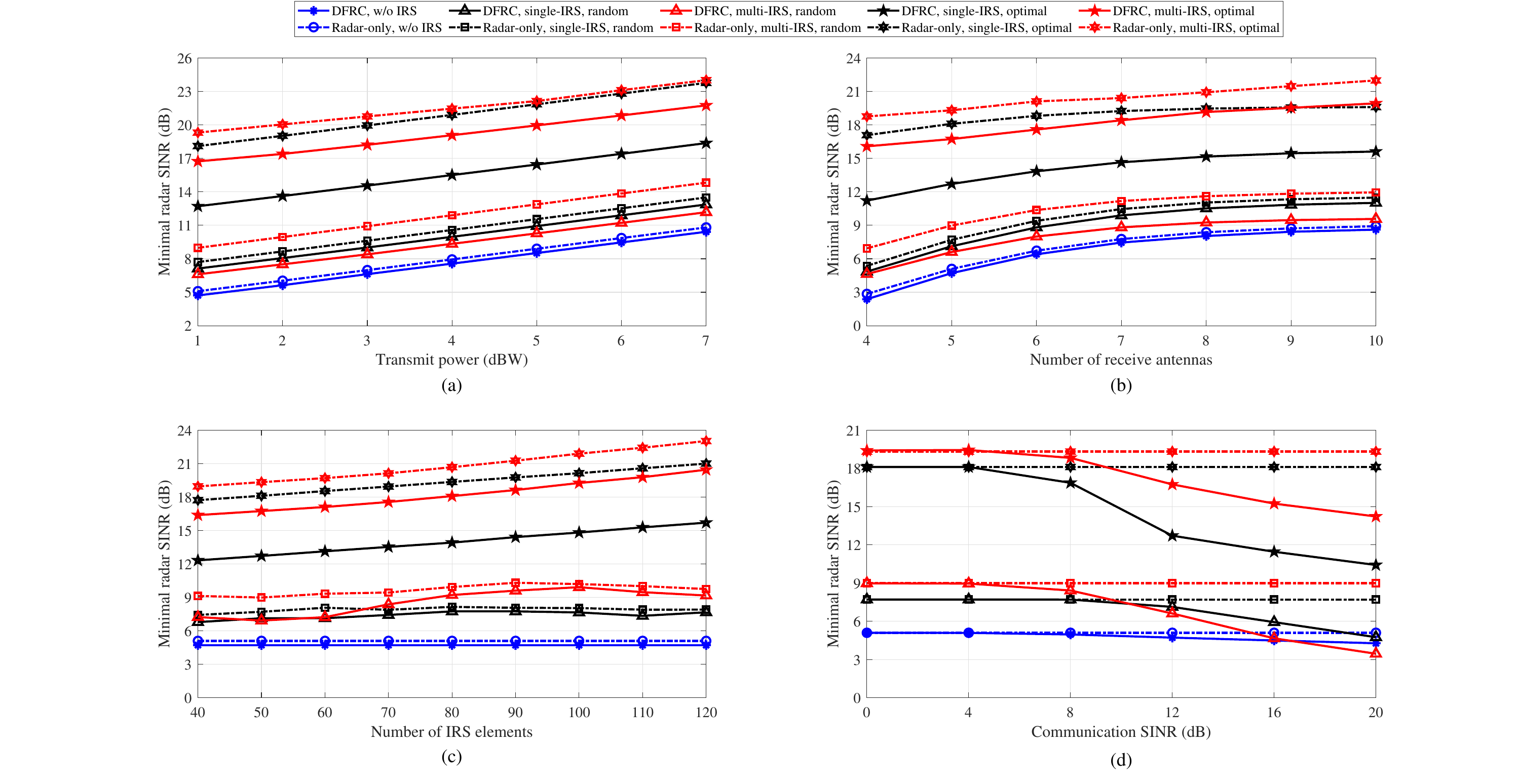}
\vspace{-0.1cm}\caption{Performance of radar SINR;  (a) Minimum radar SINR versus transmit power $\mathcal{P}_k$,   (b) Minimum radar SINR versus number of receive antennas $N_{B_r}$, (c) Minimum radar SINR versus Number of IRS elements $N_{I_m}$,   (d) Minimum radar SINR versus threshold of communications SINR $\xi$.}
\vspace{-0.1cm}\label{fig:comb}
\end{figure*}

{Fig.~\ref{fig:comb}a demonstrates the minimum radar SINR versus the transmit power of each subcarrier. The increase of the transmit power obviously improves the minimum achievable radar SINR for all systems. The single-IRS DFRC with random phase-shifts has a better performance than the multi-IRS with random phase-shift because, if the phase-shifts are not optimized, multiple IRS deployments lead to massive interference from clutter scatterers often exceeding the signal gain from the target. Furthermore, the multi-IRS DFRC with the optimal phase-shift certainly provides the best radar SINR for the different power cases of DFRC because of the higher indirect path gain. 
Fig.~\ref{fig:comb}b presents the minimum radar SINR versus the number of receive antennas. As expected, the increase in receive antennas improves the radar SINR. It yields at least $0.5$ dB enhancement per receive antenna if the receive antenna number 
$N_{B_r}$ less than 7. However, as the increasing of receive antenna, 
the SINR enhancement per receive antenna is lessened.  
Notice that in DRFC system, the proposed multi-IRS-aided system with optimal phase-shifts obtains the highest radar SINR compared with others. } 

Fig.~\ref{fig:comb}c displays the minimum radar SINR with respect to the reflecting elements of IRS. It is interesting to note that the increase in the number of IRS elements brings the obvious improvement of radar SINR for the multi-IRS DFRC with the optimal phase-shift compared with the non-IRS and single-IRS cases. For IRS-aided DFRC with random phase-shifts, an increase in reflecting elements may lead to a degradation of radar SINR, thereby directly underlining the importance of phase-shift optimization. 

{Fig.~\ref{fig:comb}d plots the minimum radar SINR versus the communications SINR. We observe that the random IRS is ineffective in DFRC system when the communications SINR greater than 16 dB. Meanwhile, multi-IRS DFRC with optimal phase-shifts always provides an enhanced radar performance even though the high communications SINR threshold is set, i.e., $\xi=20$ dB.
Compared to a standalone radar, an increase of communications SINR in DFRC leads to radar SINR loss; a higher communications SINR requirement exacerbates this deterioration in radar performance. It is, tehrefore, clear that IRS-aided DFRC system requires a performance trade-off between radar and communications necessitated by resource sharing.}

\begin{figure}[!t]
\centering
    \includegraphics[width=0.48\textwidth]{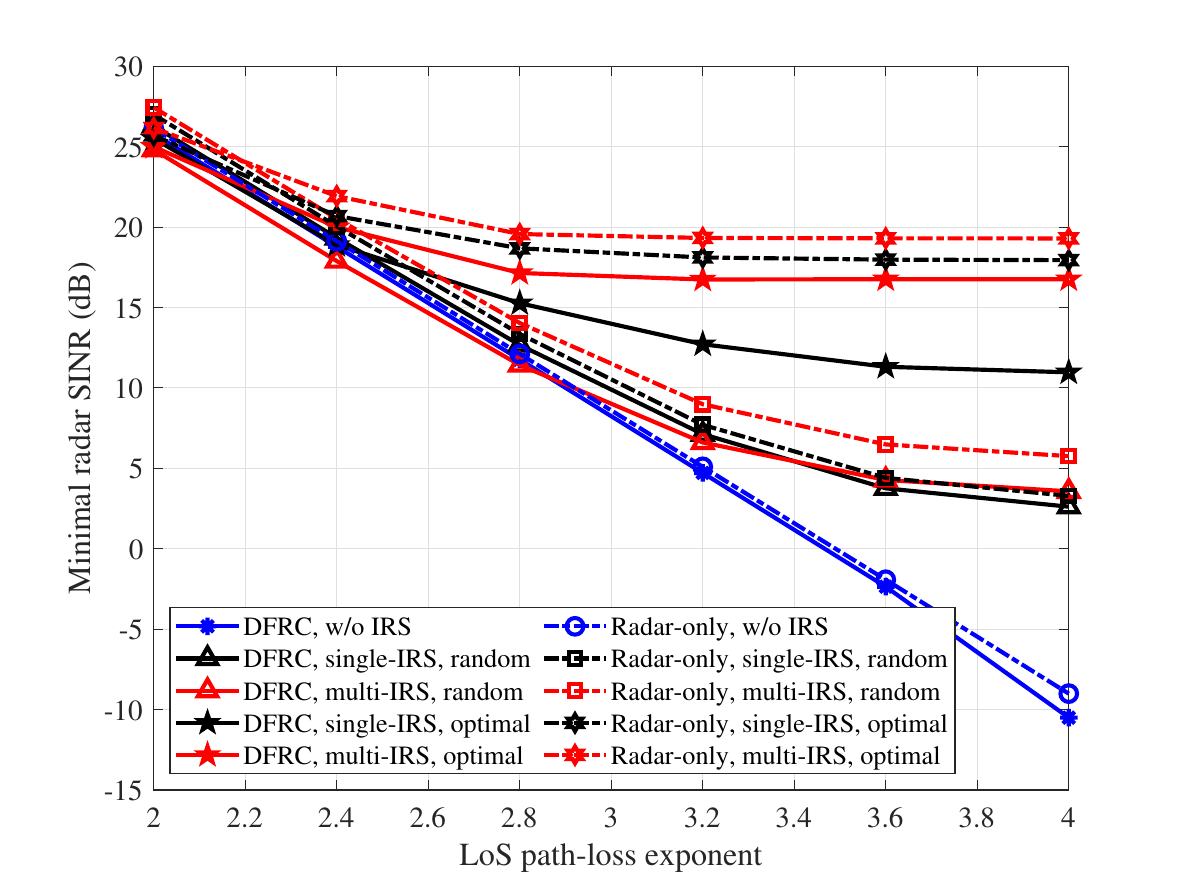}
\vspace{-0.3cm}\caption{Performance of minimum radar SINR versus LoS path loss exponent.}
\vspace{-0.5cm}\label{fig:pathloss}
\end{figure} 

{Fig.~\ref{fig:pathloss} demonstrates the minimum radar SINR versus the LoS path loss exponent. We observed that both the random IRS and optimal IRS could not provide the preferable radar SINR gain (even worsen the radar SINR in DFRC system with multiple random IRSs case) when the LoS path loss exponent is small, e.g., $\epsilon_1=2$. This means that if there exists a very strong LoS path, IRS only provides the marginal gain for the DFRC system. However, in the weak LoS path scenario, e.g., $\epsilon_1=4$,   
multi-IRS-assisted DFRC system achieves much more radar SINR gain compred with single-IRS and non-IRS system. Based on above, we conclude that multi-IRS is more effective when there does not exist a stable LoS path.   
Table~\ref{tab:performance_comparison} compares the minimum radar SINR for wideband and narrowband beamforming in different  scenarios. Note that for the narrow band beamforming, we assume a single precoder for the entire bandwidth and is implemented by using a single subcarrier, i.e.  $K=1$.  It is seen that the wideband scheme can always achieve the higher SINR gain compared with its narrowband counterpart, especially for DFRC system. This is because the fully utilized subcarrier leads to the higher degree of freedom (DoF) for beamforming design. }

\begin{table*}[t]
  \centering
  \caption{Radar SINR for wideband and narrowband beamforming}
  \vspace{-0.1cm}
  \label{tab:performance_comparison}
    \begin{tabular}{c|c|c||c|c|c}
    \hline
    Method                      &Narrowband   & Wideband  &Method                          & Narrowband    & Wideband  \cr\hline\hline
    DFRC, w/o IRS               &4.137  dB    &4.714  dB  &Radar-only, w/o IRS             &4.290 dB       &5.093 dB  \cr\hline
    DFRC, single-IRS, random    &4.365  dB    &7.125  dB  &Radar-only, single-IRS, random  &4.467 dB       &7.706 dB \cr\hline
    DFRC, multi-IRS, random     &3.842  dB    &6.609 dB   &Radar-only, multi-IRS, random   &6.196 dB       &8.985 dB\cr\hline
    DFRC, single-IRS, optimal   &10.047 dB    &12.715 dB  &Radar-only, single-IRS, optimal &15.075 dB      &18.110 dB\cr\hline 
    DFRC, multi-IRS, optimal    &11.071 dB    &16.741 dB  &Radar-only, multi-IRS, optimal  &19.098 dB      &19.323 dB\cr\hline     
    \end{tabular}    
\end{table*}

\subsection{Communications capacity and radar detection}
\begin{figure*}[!t]\centering
\vspace{-0.5cm}  \subfloat[]{
    \includegraphics[width=0.45\textwidth]{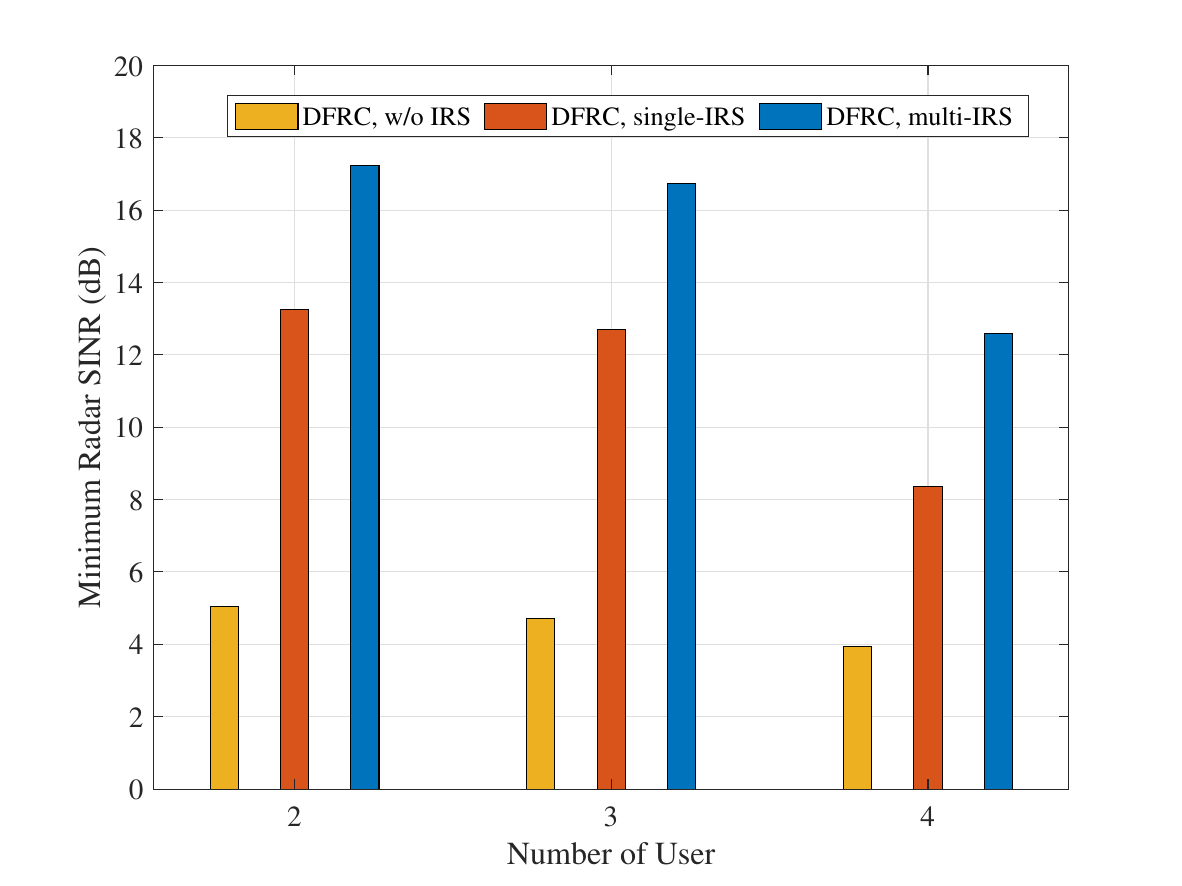}
    }
  \subfloat[]{
    \includegraphics[width=0.45\textwidth]{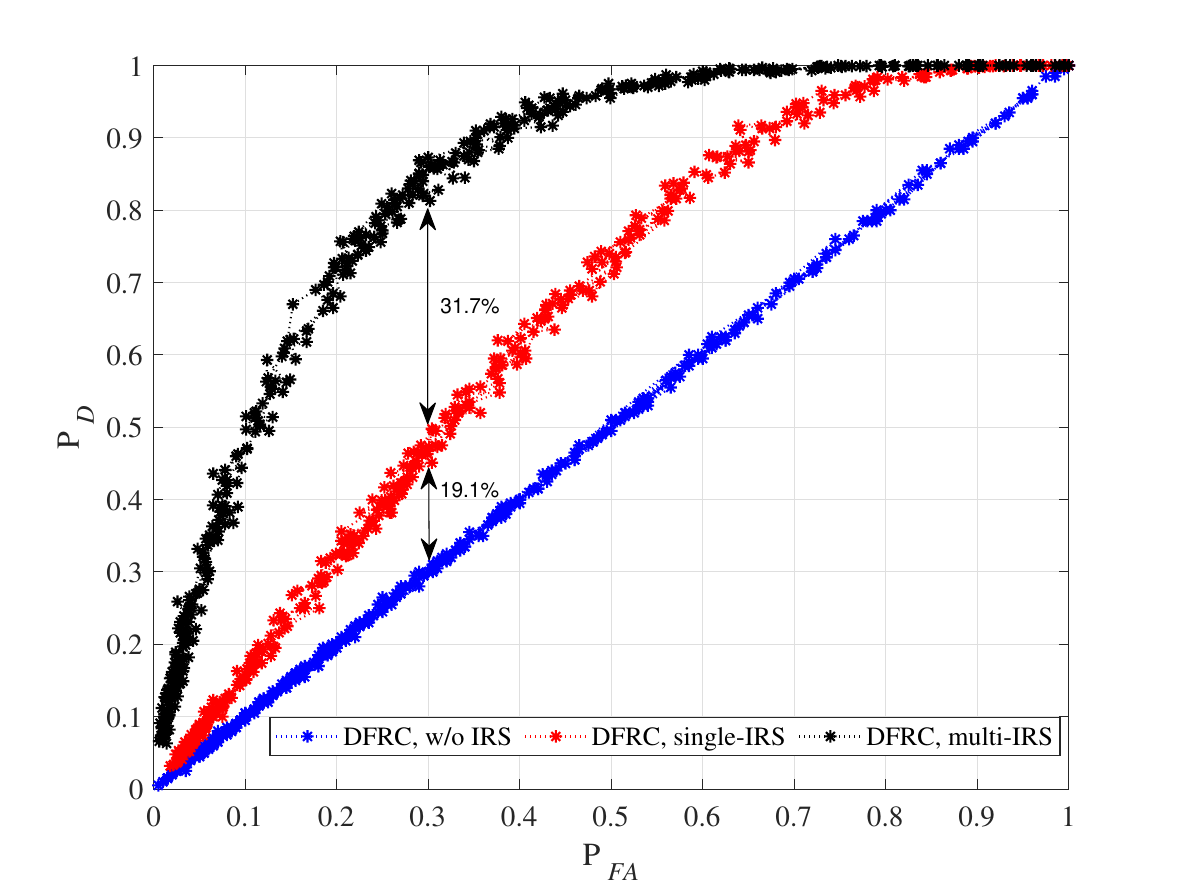}
    }  
\vspace{-0.1cm}\caption{Performance of communications  capacity and radar detection;  (a) Minimum radar SINR versus the number of users   (b) Radar detection probability versus false alarm probability.}
\vspace{-0.8cm}\label{fig:detection}
\end{figure*}

\noindent\textbf{Communications capacity:} 
{Fig.~\ref{fig:detection}a depicts the achievable worst-case radar SINR versus the number of users. It again indicates that the proposed method with the optimal multi-IRS (denoted by “DFRC, multi-IRS") achieves the best radar SINR compared with non-IRS (denoted by “DFRC, w/o IRS")  and optimal single-IRS (denoted by “DFRC, single-IRS")  cases in terms of different user numbers. On the other side, the increase of communications user can lead to the radar SINR loss for all cases. However, the proposed method can reach the preferable radar SINR regardless of the user number.}     

\noindent\textbf{Radar detection:} {The receiver operating characteristic curve that plots $P_{{D}}$ versus $P_{{FA}}$ is shown in Fig.~\ref{fig:detection}b. It is seen that with the same false alarm probability, e.g., $P_{{FA}}=0.3$, the proposed multi-IRS DFRC system achieves the highest detection probability compared with the non-IRS and single-IRS cases, effectively highlighting the superiority of our proposed multi-IRS-aided DFRC system. }

\subsection{Impact of clutter scatterers}

\begin{figure}[!t]
\centering
    \includegraphics[width=0.45\textwidth]{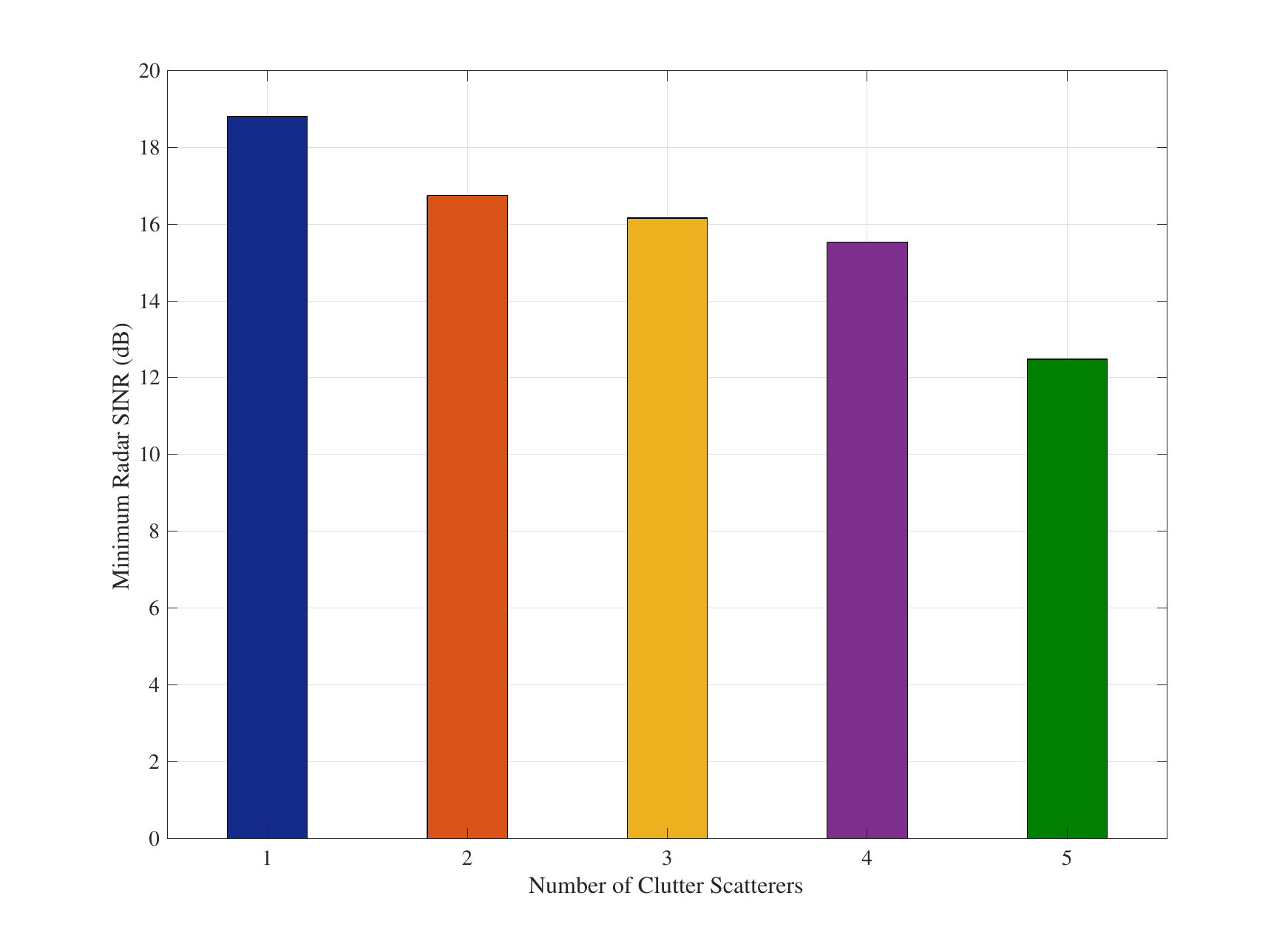}
\\\caption{Minimum radar SINR
versus the number of clutter scatterers.}
\vspace{-0.2cm}\label{fig:clutter}
\end{figure} 

{{%In general, the number and strength of clutter  scatterers affects the overall system performance compared with non-clutter systems. To express this impact more clearly, 
In Fig. \ref{fig:clutter}, we evaluated the minimum radar SINR versus the number of clutter scatterers. We considered five stationary clutter scatterers located at ${\bf p}_{C_1}=[40~\mathrm{m},38~\mathrm{m}]$, ${\bf p}_{C_2}=[100~\mathrm{m},30~\mathrm{m}]$, ${\bf p}_{C_3}=[95~\mathrm{m},65~\mathrm{m}]$, ${\bf p}_{C_4}=[90~\mathrm{m},60~\mathrm{m}]$ and ${\bf p}_{C_5}=[85~\mathrm{m},55~\mathrm{m}]$, respectively.     
Without loss of generality, the RCS for all clutters was set to $\alpha_{c_q}=1, \forall q$. We observe that the minimum radar SINR reduces with an increase in the number of clutter scatterers. With $Q=5$, radar SINR loss is $6.32$ dB when compared with $Q=1$ case.}}

\subsection{Transmit beampattern and RIS beampattern}

\begin{figure}[!t]
\centering
\includegraphics[width=0.5\textwidth]{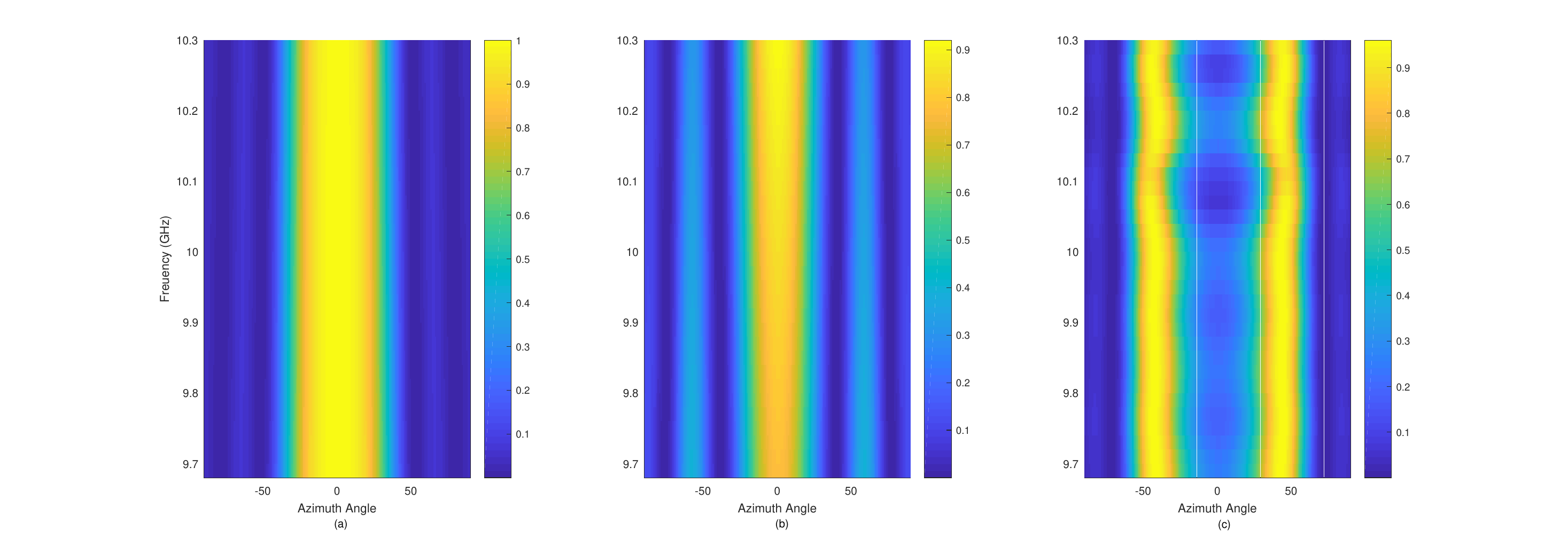}
\caption{Illustration of the normalized spatial-frequency spectrum for different scenarios, $U=2$, $M=2$, $\theta_{BT}=0^{\circ}$, $\theta_{BR_1}=45^{\circ}$, $\theta_{BR_2}=-45^{\circ}$, $\theta_{BU_1}=60^{\circ}$, and $\theta_{BU_2}=-60^{\circ}$, (a) radar-only system without IRS, (b) DFRC system without IRS, and (c) DFRC system with two IRSs but without radar LoS path.}
\label{fig:pattern}
\end{figure}

\begin{figure}[!t]
\centering
\includegraphics[width=0.5\textwidth]{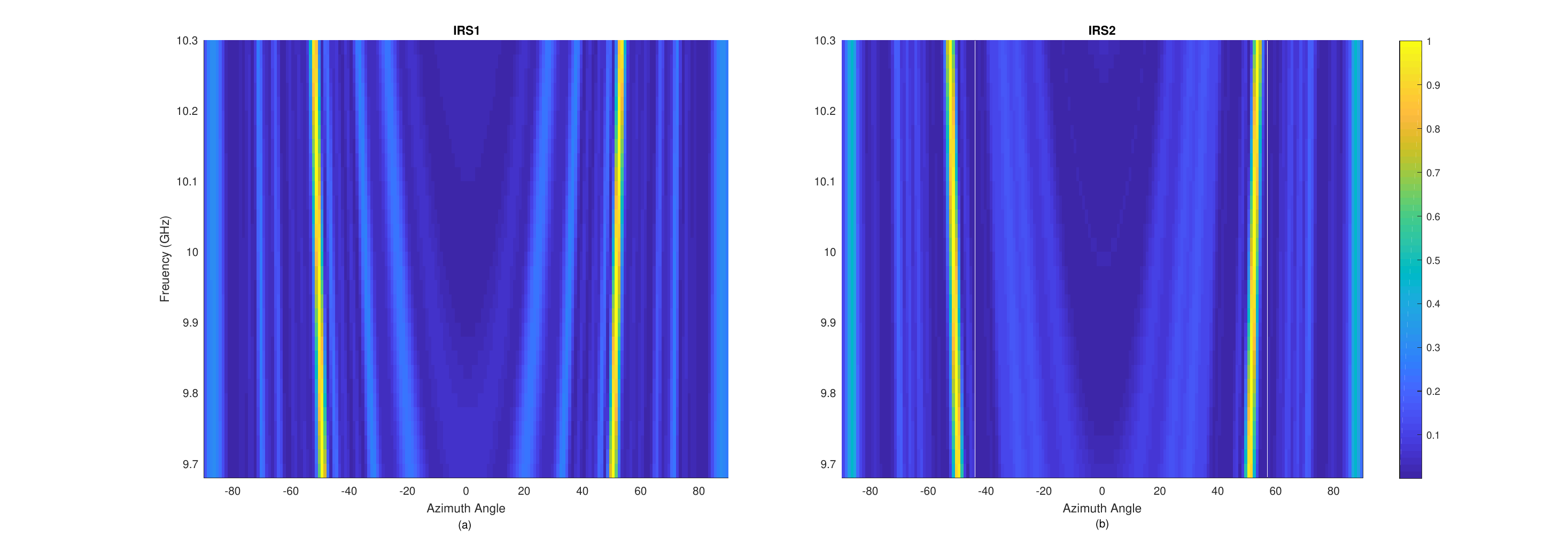}
\caption{Illustration of the normalized spatial-frequency spectrum in different IRS, (a) beampattern for IRS$_1$, $\theta_{I_1B}=-45^{\circ}$, $\theta_{I_1T}=45^{\circ}$, $\theta_{I_1U1}=-65^{\circ}$ and $\theta_{I_1U2}=-35^{\circ}$, (b) beampattern for IRS$_2$, $\theta_{I_2B}=45^{\circ}$,  $\theta_{I_2T}=-45^{\circ}$, $\theta_{I_2U1}=35^{\circ}$ and $\theta_{I_2U2}=65^{\circ}$.}
\label{fig:RIS}
\end{figure}

{Finally, we investigate the normalized spatio-spectral characteristics of BS and IRS. We consider the beampatterns for (1) BS, radar-only system without IRS, (2) BS, DFRC system without IRS, (3) BS, DFRC system with two IRSs but without LoS path, (4) IRS$_1$, DFRC system with two IRSs but without LoS path, and (5) IRS$_2$, DFRC system with two IRSs but without LoS path. Fig.~\ref{fig:pattern} and Fig.~\ref{fig:RIS} show the frequency-angle beampattern for BS and RIS, respectively. It follows from Fig.~\ref{fig:pattern}a that the transmit beam at BS is aligned to the target direction for all subcarriers in radar-only system without IRS. As for the DFRC system without IRS (Fig.~\ref{fig:pattern}b), a part of the power to the target direction is allocated to the direction of users to guarantee the communications SINR. When the LoS is not accessible (Fig.~\ref{fig:pattern}c), the BS intended to transmit two beams to cover the IRSs and users region in the DFRC system with two IRSs. From the IRS beampattern (Fig.~\ref{fig:RIS}), it follows that the IRS can form two narrow beams to the direction of BS and target; at the same time, it has two sub-beams to the direction of users. This follows because we focus on the radar SINR maximization that ensures a required communications QoS. Note that IRS here is a passive narrowband device and, hence, employs the same phase-shift for all subcarriers. As a result, beam-squint is still visible in Fig.~\ref{fig:RIS}.}

\section{Summary}
\label{sec:summ}
We proposed a novel multi-IRS-aided wideband DFRC architecture with OFDM signaling in which both the LoS and NLoS are jointly processed. Contrary to prior works, this setup is  general and includes moving target in its formulation. We focused on a radar-centric design where our goal was to obtain beamformers and IRS phase shifts that are robust to Doppler shifts of the target, while guaranteeing a certain quality of service to communication users. %Meanwhile, the worst-case optimization problem is formulated to against the unknown Doppler shift. Specifically, by properly designing the Doppler filter bank, frequency-dependent transmit beamforming and phase-shifts of multiple IRSs (i.e., passive beamforming), the minimum average radar SINR at different Doppler slice is maximized while ensuring the communications SINR of all users and transmit power constraint. 
The resulting nonconvex problem is decoupled into three subproblems and then solved them through an AM-based framework that was based on Dinkelbach, C-ADMM and RSD methods. 

Our numerical results reveal that the proposed multi-IRS-aided DFRC system achieves a much improved performance compared to the narrowband, non-IRS and single-IRS counterparts. Meanwhile, the proposed Doppler-tolerant design scheme achieves the superior minimum radar SINR over different Doppler slice, which is crucial for moving target detection.  

%The integration of IRS and DFRC achieves better detection, wider coverage, and NLoS sensing and communications. This technology is a promising tool in future beyond 5G networks, where there is a potential to deploy IRS for aerial \cite{lu2021aerial} and near-field \cite{dardari2021holographic} applications. All of these applications require investigation into Doppler-tolerant and wideband processing algorithms, toward which this work proposes several new techniques.

\ifCLASSOPTIONcaptionsoff
  \newpage
\fi

% that's all folks
%\clearpage
\bibliographystyle{IEEEtran}
\bibliography{main}

% Generated by IEEEtran.bst, version: 1.14 (2015/08/26)
\begin{thebibliography}{10}
\providecommand{\url}[1]{#1}
\csname url@samestyle\endcsname
\providecommand{\newblock}{\relax}
\providecommand{\bibinfo}[2]{#2}
\providecommand{\BIBentrySTDinterwordspacing}{\spaceskip=0pt\relax}
\providecommand{\BIBentryALTinterwordstretchfactor}{4}
\providecommand{\BIBentryALTinterwordspacing}{\spaceskip=\fontdimen2\font plus
\BIBentryALTinterwordstretchfactor\fontdimen3\font minus
  \fontdimen4\font\relax}
\providecommand{\BIBforeignlanguage}[2]{{%
\expandafter\ifx\csname l@#1\endcsname\relax
\typeout{** WARNING: IEEEtran.bst: No hyphenation pattern has been}%
\typeout{** loaded for the language `#1'. Using the pattern for}%
\typeout{** the default language instead.}%
\else
\language=\csname l@#1\endcsname
\fi
#2}}
\providecommand{\BIBdecl}{\relax}
\BIBdecl

\bibitem{tong2022multiple}
T.~Wei, L.~Wu, K.~V. Mishra, and M.~R.~B. Shankar, ``Multiple {IRS}-assisted
  wideband dual-function radar-communication,'' in \emph{IEEE International
  Symposium on Joint Communications \& Sensing}, 2022, pp. 1--5.

\bibitem{renzo2020smart}
M.~Di~Renzo, A.~Zappone, M.~Debbah, M.-S. Alouini, C.~Yuen, J.~de~Rosny, and
  S.~Tretyakov, ``Smart radio environments empowered by reconfigurable
  intelligent surfaces: {How} it works, state of research, and the road
  ahead,'' \emph{IEEE Journal on Selected Areas in Communications}, vol.~38,
  no.~11, pp. 2450--2525, 2020.

\bibitem{huang2019reconfigurable}
C.~Huang \emph{et~al.}, ``Reconfigurable intelligent surfaces for energy
  efficiency in wireless communication,'' \emph{IEEE Transactions on Wireless
  Communications}, vol.~18, no.~8, pp. 4157--4170, 2019.

\bibitem{wu2020towards}
Q.~Wu and R.~Zhang, ``Towards smart and reconfigurable environment:
  {I}ntelligent reflecting surface aided wireless network,'' \emph{IEEE
  Communications Magazine}, vol.~58, no.~1, pp. 106--112, 2020.

\bibitem{hodge2020intelligent}
J.~A. Hodge, K.~V. Mishra, and A.~I. Zaghloul, ``Intelligent time-varying
  metasurface transceiver for index modulation in {6G} wireless networks,''
  \emph{IEEE Antennas and Wireless Propagation Letters}, vol.~19, no.~11, pp.
  1891--1895, 2020.

\bibitem{garcia2020reconfigurable}
J.~C.~B. Garcia, A.~Sibille, and M.~Kamoun, ``Reconfigurable intelligent
  surfaces: {Bridging} the gap between scattering and reflection,'' \emph{IEEE
  Journal on Selected Areas in Communications}, vol.~38, no.~11, pp.
  2538--2547, 2020.

\bibitem{aubry2021reconfigurable}
A.~Aubry, A.~De~Maio, and M.~Rosamilia, ``Reconfigurable intelligent surfaces
  for {N-LOS} radar surveillance,'' \emph{IEEE Transactions on Vehicular
  Technology}, vol.~70, no.~10, pp. 10\,735--10\,749, 2021.

\bibitem{zahra2021irs-aided}
Z.~Esmaeilbeig, K.~V. Mishra, and M.~Soltanalian, ``{IRS}-aided radar:
  {E}nhanced target parameter estimation via intelligent reflecting surfaces,''
  in \emph{IEEE Sensor Array and Multichannel Signal Processing Workshop},
  2021, pp. 1--5.

\bibitem{lu2021intelligent}
W.~Lu, B.~Deng, Q.~Fang, X.~Wen, and S.~Peng, ``Intelligent reflecting
  surface-enhanced target detection in {MIMO} radar,'' \emph{IEEE Sensors
  Letters}, vol.~5, no.~2, pp. 1--4, 2021.

\bibitem{xu2020resource}
D.~Xu, X.~Yu, Y.~Sun, D.~W.~K. Ng, and R.~Schober, ``Resource allocation for
  {IRS}-assisted full-duplex cognitive radio systems,'' \emph{IEEE Transactions
  on Communications}, vol.~68, no.~12, pp. 7376--7394, 2020.

\bibitem{tang2021wireless}
W.~Tang, M.~Z. Chen, X.~Chen, J.~Y. Dai, Y.~Han, M.~Di~Renzo, Y.~Zeng, S.~Jin,
  Q.~Cheng, and T.~J. Cui, ``Wireless communications with reconfigurable
  intelligent surface: {P}ath loss modeling and experimental measurement,''
  \emph{IEEE Transactions on Wireless Communications}, vol.~20, no.~1, pp.
  421--439, 2021.

\bibitem{Ur2021joint}
H.~Ur~Rehman, F.~Bellili, A.~Mezghani, and E.~Hossain, ``Joint active and
  passive beamforming design for {IRS}-assisted multi-user {MIMO} systems: {A}
  {VAMP}-based approach,'' \emph{IEEE Transactions on Communications}, vol.~69,
  no.~10, pp. 6734--6749, 2021.

\bibitem{wu2020joint}
Q.~Wu and R.~Zhang, ``Joint active and passive beamforming optimization for
  intelligent reflecting surface assisted {SWIPT} under {QoS} constraints,''
  \emph{IEEE Journal on Selected Areas in Communications}, vol.~38, no.~8, pp.
  1735--1748, 2020.

\bibitem{zhou2021joint}
Z.~Zhou, N.~Ge, Z.~Wang, and L.~Hanzo, ``Joint transmit precoding and
  reconfigurable intelligent surface phase adjustment: {A} decomposition-aided
  channel estimation approach,'' \emph{IEEE Transactions on Communications},
  vol.~69, no.~2, pp. 1228--1243, 2021.

\bibitem{abeywickrama2020intelligent}
S.~Abeywickrama, R.~Zhang, Q.~Wu, and C.~Yuen, ``Intelligent reflecting
  surface: {Practical} phase shift model and beamforming optimization,''
  \emph{IEEE Transactions on Communications}, vol.~68, no.~9, pp. 5849--5863,
  2020.

\bibitem{elzanaty2021reconfigurable}
A.~Elzanaty, A.~Guerra, F.~Guidi, and M.-S. Alouini, ``Reconfigurable
  intelligent surfaces for localization: {Position} and orientation error
  bounds,'' \emph{IEEE Transactions on Signal Processing}, vol.~69, pp.
  5386--5402, 2021.

\bibitem{yu2020robust}
X.~Yu, D.~Xu, Y.~Sun, D.~W.~K. Ng, and R.~Schober, ``Robust and secure wireless
  communications via intelligent reflecting surfaces,'' \emph{IEEE Journal on
  Selected Areas in Communications}, vol.~38, no.~11, pp. 2637--2652, 2020.

\bibitem{zheng2021double}
B.~Zheng, C.~You, and R.~Zhang, ``Double-{IRS} assisted multi-user {MIMO}:
  {Cooperative} passive beamforming design,'' \emph{IEEE Transactions on
  Wireless Communications}, vol.~20, no.~7, pp. 4513--4526, 2021.

\bibitem{mei2021cooperative}
W.~Mei and R.~Zhang, ``Cooperative beam routing for multi-{IRS} aided
  communication,'' \emph{IEEE Wireless Communications Letters}, vol.~10, no.~2,
  pp. 426--430, 2021.

\bibitem{he2022ris}
Y.~He, Y.~Cai, H.~Mao, and G.~Yu, ``{RIS}-assisted communication radar
  coexistence: Joint beamforming design and analysis,'' \emph{IEEE Journal on
  Selected Areas in Communications}, 2022, in press.

\bibitem{jiang2022intelligent}
Z.-M. Jiang, M.~Rihan, P.~Zhang, L.~Huang, Q.~Deng, J.~Zhang, and E.~M.
  Mohamed, ``Intelligent reflecting surface aided dual-function radar and
  communication system,'' \emph{IEEE Systems Journal}, vol.~16, no.~1, pp.
  475--486, 2022.

\bibitem{wang2021ris}
X.~Wang, Z.~Fei, J.~Guo, Z.~Zheng, and B.~Li, ``{RIS}-assisted spectrum sharing
  between {MIMO} radar and {MU-MISO} communication systems,'' \emph{IEEE
  Wireless Communications Letters}, vol.~10, no.~3, pp. 594--598, 2021.

\bibitem{liu2022joint}
R.~Liu, M.~Li, Y.~Liu, Q.~Wu, and Q.~Liu, ``Joint transmit waveform and passive
  beamforming design for {RIS}-aided {DFRC} systems,'' \emph{IEEE Journal of
  Selected Topics in Signal Processing}, pp. 1--15, 2022.

\bibitem{song2022joint}
X.~Song, D.~Zhao, H.~Hua, T.~X. Han, X.~Yang, and J.~Xu, ``Joint transmit and
  reflective beamforming for {IRS-Assisted} integrated sensing and
  communication,'' in \emph{IEEE Wireless Communications and Networking
  Conference}, 2022, pp. 189--194.

\bibitem{zhu2022intelligent}
Z.~Zhu, Z.~Li, Z.~Chu, G.~Sun, W.~Hao, P.~Xiao, and I.~Lee, ``Intelligent
  reflecting surface assisted integrated sensing and communications for
  {mmWave} channels,'' \emph{arXiv preprint arXiv:2202.00552}, 2022.

\bibitem{hua2022joint}
M.~Hua, Q.~Wu, C.~He, S.~Ma, and W.~Chen, ``Joint active and passive
  beamforming design for {IRS}-aided radar-communication,'' \emph{arXiv
  preprint arXiv:2203.14532}, 2022.

\bibitem{wang2021joint}
X.~Wang, Z.~Fei, Z.~Zheng, and J.~Guo, ``Joint waveform design and passive
  beamforming for {RIS}-assisted dual-functional radar-communication system,''
  \emph{IEEE Transactions on Vehicular Technology}, vol.~70, no.~5, pp.
  5131--5136, 2021.

\bibitem{wang2022jointwaveform}
X.~Wang, Z.~Fei, J.~Huang, and H.~Yu, ``Joint waveform and discrete phase shift
  design for {RIS}-assisted integrated sensing and communication system under
  cram\'{e}r-{R}ao bound constraint,'' \emph{IEEE Transactions on Vehicular
  Technology}, vol.~71, no.~1, pp. 1004--1009, 2022.

\bibitem{esmaeilbeig2023quantized}
Z.~Esmaeilbeig, A.~Eamaz, K.~V. Mishra, and M.~Soltanalian, ``Quantized
  phase-shift design of active {IRS} for integrated sensing and
  communications,'' in \emph{IEEE International Conference on Acoustics,
  Speech, \& Signal Processing Workshops}, 2023, in press.

\bibitem{buzzi2022foundations}
S.~Buzzi, E.~Grossi, M.~Lops, and L.~Venturino, ``Foundations of {MIMO} radar
  detection aided by reconfigurable intelligent surfaces,'' \emph{IEEE
  Transactions on Signal Processing}, pp. 1--1, 2022.

\bibitem{wang2022joint}
F.~Wang, H.~Li, and J.~Fang, ``Joint active and passive beamforming for
  {IRS}-assisted radar,'' \emph{IEEE Signal Processing Letters}, vol.~29, pp.
  349--353, 2022.

\bibitem{aubry2021ris}
A.~Aubry, A.~De~Maio, and M.~Rosamilia, ``{RIS}-aided radar sensing in {N-LOS}
  environment,'' in \emph{IEEE International Workshop on Metrology for
  Aerospace}, 2021, pp. 277--282.

\bibitem{watson2019non}
B.~Watson and J.~R. Guerci, \emph{Non-line-of-sight radar}.\hskip 1em plus
  0.5em minus 0.4em\relax Artech House, 2019.

\bibitem{wei2022nonline}
S.~Wei, J.~Wei, X.~Liu, M.~Wang, S.~Liu, F.~Fan, X.~Zhang, J.~Shi, and G.~Cui,
  ``Nonline-of-sight {3-D} imaging using millimeter-wave radar,'' \emph{IEEE
  Transactions on Geoscience and Remote Sensing}, vol.~60, pp. 1--18, 2022.

\bibitem{sankar2021joint}
R.~S. Prasobh~Sankar, B.~Deepak, and S.~P. Chepuri, ``Joint communication and
  radar sensing with reconfigurable intelligent surfaces,'' in \emph{IEEE
  International Workshop on Signal Processing Advances in Wireless
  Communications}, 2021, pp. 471--475.

\bibitem{sankar2022beamforming}
\BIBentryALTinterwordspacing
R.~S.~P. Sankar, S.~P. Chepuri, and Y.~C. Eldar, ``Beamforming in integrated
  sensing and communication systems with reconfigurable intelligent surfaces,''
  2022. [Online]. Available: \url{https://arxiv.org/abs/2206.07679}
\BIBentrySTDinterwordspacing

\bibitem{liu2022joint2}
R.~Liu, M.~Li, and A.~L. Swindlehurst, ``Joint beamforming and reflection
  design for {RIS}-assisted isac systems,'' in \emph{2022 30th European Signal
  Processing Conference (EUSIPCO)}, 2022, pp. 997--1001.

\bibitem{mishra2019toward}
K.~V. Mishra, M.~R.~B. Shankar, V.~Koivunen, B.~Ottersten, and S.~A. Vorobyov,
  ``Toward millimeter-wave joint radar communications: {A} signal processing
  perspective,'' \emph{IEEE Signal Processing Magazine}, vol.~36, no.~5, pp.
  100--114, 2019.

\bibitem{wu2022resource}
L.~Wu, K.~V. Mishra, M.~R.~B. Shankar, and B.~Ottersten, ``Resource allocation
  in heterogeneously-distributed joint radar-communications under asynchronous
  {B}ayesian tracking framework,'' \emph{IEEE Journal on Selected Areas in
  Communications}, vol.~40, no.~7, pp. 2026--2042, 2022.

\bibitem{liu2020joint}
F.~Liu, C.~Masouros, A.~P. Petropulu, H.~Griffiths, and L.~Hanzo, ``Joint radar
  and communication design: {A}pplications, state-of-the-art, and the road
  ahead,'' \emph{IEEE Transactions on Communications}, vol.~68, no.~6, pp.
  3834--3862, 2020.

\bibitem{luo2022joint}
H.~Luo, R.~Liu, M.~Li, Y.~Liu, and Q.~Liu, ``Joint beamforming design for
  {RIS}-assisted integrated sensing and communication systems,'' \emph{IEEE
  Transactions on Vehicular Technology}, vol.~71, no.~12, pp. 13\,393--13\,397,
  2022.

\bibitem{hassanien2016dual}
A.~Hassanien \emph{et~al.}, ``Dual-function radar-communications: Information
  embedding using sidelobe control and waveform diversity,'' \emph{IEEE
  Transactions on Signal Processing}, vol.~64, no.~8, pp. 2168--2181, 2016.

\bibitem{irs_jrc_SecrecyRateOptMishra}
K.~V. Mishra, A.~Chattopadhyay, S.~S. Acharjee, and A.~P. Petropulu,
  ``{OptM3Sec}: {O}ptimizing multicast {IRS}-aided multiantenna {DFRC} secrecy
  channel with multiple eavesdroppers,'' in \emph{IEEE International Conference
  on Acoustics, Speech and Signal Processing}, 2022, pp. 9037--9041.

\bibitem{liu2010wideband}
W.~Liu and S.~Weiss, \emph{Wideband beamforming: {Concepts} and
  techniques}.\hskip 1em plus 0.5em minus 0.4em\relax John Wiley \& Sons, 2010.

\bibitem{ma2021wideband}
S.~Ma, W.~Shen, J.~An, and L.~Hanzo, ``Wideband channel estimation for
  {IRS}-aided systems in the face of beam squint,'' \emph{IEEE Transactions on
  Wireless Communications}, vol.~20, no.~10, pp. 6240--6253, 2021.

\bibitem{cheng2021transmit}
Z.~Cheng, S.~Shi, Z.~He, and B.~Liao, ``Transmit sequence design for
  dual-function radar-communication system with one-bit {DACs},'' \emph{IEEE
  Transactions on Wireless Communications}, vol.~20, no.~9, pp. 5846--5860,
  2021.

\bibitem{cheng2021hybrid}
Z.~Cheng, Z.~He, and B.~Liao, ``Hybrid beamforming design for {OFDM}
  dual-function radar-communication system,'' \emph{IEEE Journal of Selected
  Topics in Signal Processing}, vol.~15, no.~6, pp. 1455--1467, 2021.

\bibitem{sun2019target}
S.~Sun, K.~V. Mishra, and A.~P. Petropulu, ``Target estimation by exploiting
  low rank structure in widely separated {MIMO} radar,'' in \emph{IEEE Radar
  Conference}, 2019, pp. 1--6.

\bibitem{he2021channel}
J.~He, H.~Wymeersch, and M.~Juntti, ``Channel estimation for {RIS}-aided
  {mmWave} {MIMO} systems via atomic norm minimization,'' \emph{IEEE
  Transactions on Wireless Communications}, vol.~20, no.~9, pp. 5786--5797,
  2021.

\bibitem{aubry2015optimizing}
A.~Aubry, A.~De~Maio, and M.~M. Naghsh, ``Optimizing radar waveform and
  {Doppler} filter bank via generalized fractional programming,'' \emph{IEEE
  Journal of Selected Topics in Signal Processing}, vol.~9, no.~8, pp.
  1387--1399, 2015.

\bibitem{tsinos2021joint}
C.~G. Tsinos, A.~Arora, S.~Chatzinotas, and B.~Ottersten, ``Joint transmit
  waveform and receive filter design for dual-function radar-communication
  systems,'' \emph{IEEE Journal of Selected Topics in Signal Processing},
  vol.~15, no.~6, pp. 1378--1392, 2021.

\bibitem{park2017dynamic}
S.~Park, A.~Alkhateeb, and R.~W. Heath, ``Dynamic subarrays for hybrid
  precoding in wideband mmwave {MIMO} systems,'' \emph{IEEE Transactions on
  Wireless Communications}, vol.~16, no.~5, pp. 2907--2920, 2017.

\bibitem{wei2021channel}
L.~Wei, C.~Huang, G.~C. Alexandropoulos, C.~Yuen, Z.~Zhang, and M.~Debbah,
  ``Channel estimation for {RIS}-empowered multi-user {MISO} wireless
  communications,'' \emph{IEEE Transactions on Communications}, vol.~69, no.~6,
  pp. 4144--4157, 2021.

\bibitem{bertsekas1997nonlinear}
D.~P. Bertsekas, ``Nonlinear programming,'' \emph{Journal of the Operational
  Research Society}, vol.~48, no.~3, pp. 334--334, 1997.

\bibitem{charnes1968programming}
A.~Charnes and W.~W. Cooper, ``Programming with linear fractional
  functionals,'' \emph{Naval Research Logistics Quarterly}, vol.~9, no. 3-4,
  pp. 181--186, 1962.

\bibitem{huang2010rank}
Y.~Huang and D.~P. Palomar, ``Rank-constrained separable semidefinite
  programming with applications to optimal beamforming,'' \emph{IEEE
  Transactions on Signal Processing}, vol.~58, no.~2, pp. 664--678, 2010.

\bibitem{luo2010semidefinite}
Z.-Q. Luo, W.-K. Ma, A.~M.-C. So, Y.~Ye, and S.~Zhang, ``Semidefinite
  relaxation of quadratic optimization problems,'' \emph{IEEE Signal Processing
  Magazine}, vol.~27, no.~3, pp. 20--34, 2010.

\bibitem{petersen2008matrix}
K.~B. Petersen and M.~S. Pedersen, \emph{The matrix cookbook}.\hskip 1em plus
  0.5em minus 0.4em\relax Technical University of Denmark, 2012.

\bibitem{zhang2021joint}
Z.~Zhang and L.~Dai, ``A joint precoding framework for wideband reconfigurable
  intelligent surface-aided cell-free network,'' \emph{IEEE Transactions on
  Signal Processing}, vol.~69, pp. 4085--4101, 2021.

\bibitem{yang2020dual}
J.~Yang, G.~Cui, X.~Yu, and L.~Kong, ``Dual-use signal design for radar and
  communication via ambiguity function sidelobe control,'' \emph{IEEE
  Transactions on Vehicular Technology}, vol.~69, no.~9, pp. 9781--9794, 2020.

\bibitem{grant2009cvx}
M.~Grant, S.~Boyd, and Y.~Ye, ``{CVX}: {MATLAB} software for disciplined convex
  programming,'' 2009.

\bibitem{duggal2020doppler}
G.~Duggal, S.~Vishwakarma, K.~V. Mishra, and S.~S. Ram, ``Doppler-resilient
  802.11ad-based ultrashort range automotive joint radar-communications
  system,'' \emph{IEEE Transactions on Aerospace and Electronic Systems},
  vol.~56, no.~5, pp. 4035--4048, 2020.

\end{thebibliography}

%\begin{IEEEbiography}[{\includegraphics[width=1in,height=1.25in,clip,keepaspectratio]{Tong_wei.jpg}}]{Tong Wei} (Student Member, IEEE) received
%the B.E. degree from Hainan University, Hainan, China, and M.Sc. degree from Shenzhen University, Shenzhen, China, in 2017 and 2020, respectively.  
%He is currently working toward the Ph.D. degree with the Signal Processing
%Applications in Radar and Communications (SPARC) Group at the
%University of Luxembourg, Luxembourg City, Luxembourg. His research
%interests include array signal processing, detection $\&$ estimation theory, optimization with
%applications to wideband waveform design, integrated sensing and communication, and reconfigurable metasurfaces. \end{IEEEbiography}

\ifCLASSOPTIONcaptionsoff
  \newpage
\fi

\end{document}